\documentclass[acmsmall]{acmart}
\AtBeginDocument{%
  \providecommand\BibTeX{{%
    \normalfont B\kern-0.5em{\scshape i\kern-0.25em b}\kern-0.8em\TeX}}}

\setcopyright{acmcopyright}
\copyrightyear{2024}
\acmYear{2024}
\acmDOI{XXXXXXX.XXXXXXX}

\acmPrice{15.00}
\acmISBN{978-1-4503-XXXX-X/18/06}

\usepackage{hyperref}
\usepackage{url}
\usepackage[utf8]{inputenc}
\usepackage{subfigure}
\usepackage{caption}
\usepackage{subcaption}
\usepackage{booktabs}
\usepackage{graphicx}
\usepackage{colortbl}
\usepackage{lscape}
\begin{document}

\title{Why Larp?! A Synthesis Paper on Live Action Roleplay in Relation to HCI Research and Practice}


\author{Karin Johansson}
\affiliation{%
  \institution{Uppsala University}
  \country{Sweden}}
  \email{karin.b.johansson@im.uu.se}
  
\author{Raquel Robinson}
\affiliation{%
  \institution{IT University of Copenhagen}
  \country{Denmark}}
  \email{raqr@itu.dk}

  \author{Jon Back}
\affiliation{%
  \institution{Uppsala University}
  \country{Sweden}}
  \email{jon.back@im.uu.se }

    \author{Sarah Lynne Bowman}
\affiliation{%
  \institution{Uppsala University}
  \country{Sweden}}
  \email{sarah.bowman@speldesign.uu.se}

  \author{James Fey}
\affiliation{%
  \institution{University of California, Santa Cruz}
  \country{USA}}
  \email{jfey@ucsc.edu}
  
\author{Elena Marquez-Segura}
\affiliation{%
  \institution{Universidad Carlos III de Madrid}
  \country{Spain}}
  \email{emarquez@inf.uc3m.es}
  
    \author{Annika Waern}
\affiliation{%
  \institution{Uppsala University}
  \country{Sweden}}
  \email{annika.waern@im.uu.se}

  \author{Katherine Isbister}
\affiliation{%
  \institution{University of California, Santa Cruz}
  \country{USA}}
  \email{kisbiste@ucsc.edu}

\renewcommand{\shortauthors}{Johansson et al.}

\begin{abstract}
Live action roleplay (larp) has a wide range of applications, and can be relevant in relation to HCI. While there has been research about larp in relation to topics such as embodied interaction, playfulness and futuring published in HCI venues since the early 2000s, there is not yet a compilation of this knowledge. In this paper, we synthesise knowledge about larp and larp-adjacent work within the domain of HCI. We present a practitioner overview from an expert group of larp researchers, the results of a literature review, and highlight particular larp research exemplars which all work together to showcase the diverse set of ways that larp can be utilised in relation to HCI topics and research. This paper identifies the need for further discussions toward establishing best practices for utilising larp in relation to HCI research, as well as advocating for increased engagement with larps outside academia.

\end{abstract}





\maketitle

\section{Introduction}
Live Action Roleplay (larp) contains many elements that make it interesting in the context of Human-Computer Interaction (HCI), such as use of technologies, futuring through first-person perspectives, interaction and activity design solutions and game topics. While similar phenomenon have been around for a long time, larp in its modern form is a cultural expression and a subculture that has been around since the 1980s. In the beginnings, it mostly focused on fantasy settings \cite{Tychsen2006LiveSimilarities}. The early focus on fantasy in larp might be a contributing factor to the misconception that larp in its essence is mostly non-technological, but despite being an embodied and physical experience, larp explores technologies and futuring in an abundance of ways, and can offer powerful methods for this. In this paper we identify how larp can be relevant in relation to HCI. Larpers have been early adopters of technology, and larp also has been a way of exploring historical, current and future technologies. Larp can be seen as a type of media, a cultural expression and even a tool for learning \cite{kurz_learning_2015}. In larp, a group of people take on roles and engage in embodied co-creative storytelling creating an immersive experience together, often enhanced by clothing, props and immersive venues \cite{Salen2003}. Within this basic foundation, the purpose, number of participants, themes, type of venues, duration and so on vary greatly between different larps \cite{Tychsen2006LiveSimilarities}.

While larp has increasingly been brought to attention within the field of HCI, there is a lack of coordination and knowledge sharing between different initiatives, further there is a gap between larp related research in HCI and the vast larp practice. This paper makes a key contribution as a synthesising paper, by bringing together and combining knowledge from ongoing initiatives within and outside academia, to untangle what larp can be in relation to HCI relevant topics, finding more opportunities and ways for larp to become relevant within HCI research and practice. 

This paper brings together a group of experts who are all HCI researchers engaged with larp related research in HCI relevant topics, and who are also active larpers and/or larp designers. They are all co-authors of this paper. The knowledge within the group was brought together through extensive discussions, combined with gathering external information from written resources, from practitioners and within HCI research community. Through charting the increased use of larp in and outside of HCI, we aim to answer the following questions:

\begin{itemize}
    \item How does larp as method, practice and cultural expression relate to current topics and practices in HCI?
    \item How has larp been utilised within the HCI community?
    \item What are the virtues and challenges of using larp in relation to HCI?

\end{itemize}

For many years, the HCI community has engaged with topics such as games and play, simulations, embodied interaction, public play, augmented reality, roleplaying, and other related concepts that to a lesser or greater extent connect to larp. During the last 20 years, larp as a phenomenon in itself, and its uses for the HCI research community, has gained increasing attention. In 2019 at the Designing Interactive Systems (DIS) conference, a workshop on larp as an embodied design research method was held by a subgroup of the authors of this paper \cite{MarquezSegura2019LarpingMethod}. This sparked a lot of discussions and showed that there was an interest in larp within the HCI community. In 2023, the authors of this paper held a workshop at the premier CHI Conference on Human Factors in Computing Systems on educational larp \cite{robinson_edu-larp_2023}, which cemented the need to continue the discourse of unpacking larp and its relationship to HCI. This paper addresses this need by synthesising previous work in the field through first, a practitioner overview of larp in section \ref{PO}, followed by a more structured literature review within the ACM digital library in section \ref{ACM}. Next, in section \ref{EXE}, we present four larp research exemplars of our own work which show a broad range of how larp can be used in HCI. Lastly, we map the reviewed literature to broader themes and create classifications which highlight a number of research gaps in section \ref{discussion}. Through an extensive review of current practices and use of larp in HCI, the article argues that larp is more than a relevant domain of study. We identify unique opportunities that larps bring to HCI related to its capabilities of supporting sensitising, ideation, and exploring possible futures. 

\section{Background}
This section serves as a brief introduction to larp, defining what we see as larp and larp-related activities. This background is brief, as in later sections there will be a more comprehensive overview of larp within the context of HCI as part of our contribution.

\subsection{History of Larp}
The cultural phenomenon of larp as we know it today emerged during the early 1980s, based in tabletop roleplaying, with many of the early larps set in fantasy worlds \cite{Tychsen2006LiveSimilarities}, and, especially in the Nordic countries, also based in improvisational theater \cite{2023WhatLarp}. But there is also a long history of larp-like events before modern times, from ancient Egypt \cite{Stenros2007PervasiveSociety}, China \cite{contributor_roleplaying_2021, absurd}, Rome \cite{noauthorriddarleknodate} and more, where people dressed up, found immersive locations and took on roles. From very early on, larp-like activities became a way to also playfully investigate technological possibilities; One of the world's first robots, Da Vinci's robotic armour from 1495, seems to have been built for a divertissement, a larp-like form of entertainment \cite{Rosheim2006LeonardosRobots}, and in 1779 an advanced giant fire breathing dragon was built for a roleplay adventure arranged by the Swedish king \footnote{Karusell vid Drottningholm 1779, \url{https://digitaltmuseum.se/021046500625/karusell-vid-drottningholm-1779-erovringen-av-galtarklippan}}. 

Larp as a subculture in its modern form grew during the 1980s, initially heavily leaning on fantasy as a theme \cite{Tychsen2006LiveSimilarities}, but quickly the variations of this versatile media proliferated and genres such as science fiction and dystopic futuristic themes emerged \footnote{History of the IFGS \url{http://www.ifgs.org/history.asp}}.

\subsection{Larp Today}
Today, larp is a vibrant subculture. A conservative estimate would be that at least 13 million people participate in larps yearly \cite{xiong_player_2023}, and larps are arranged in many parts of the world; for instance China, \cite{live_nodate}, the USA \cite{DyversHandsindividualauthors2014ONECENSUS}, Europe such as \textit{Zero}, (see 23 in Appendix \ref{appendix: larps}); Nordic Larp in France \footnote{\url{https://nordiclarp.org/tag/france/}}; \textit{Mythodea}, (see 15 in Appendix \ref{appendix: larps}) \cite{Back2014TheLarp}, South Africa \cite{Robnson2017BarbariansPark}, Brazil \cite{Karlsson2014NewIce}, Japan \cite{kamm_role-playing_2020}, and Australia \cite{Tychsen2006LiveSimilarities}.

Larp is a sub-culture, but has also received attention in the wider society, through museum exhibitions \footnote{The Game of the World, \url{https://www.varldskulturmuseet.se/utstallningar/varldens-spel/}} and traditional media \cite{docu, Clasper2017PlayClassrooms}. In the experience industry, big corporations such as Disney have given attention to the design possibilities of larp  \cite{Jahromi2022LARPingWorld, Hon2023StarStarcruiser}. Larp-like methods are also used in different educational contexts, then often referred to as edu-larp \footnote{LajvVerkstaden, \url{https://lajvverkstaden.se/}} \cite{Schrier2019LearningGames,Bowman2014EducationalReview}. Outside HCI there is an abundance of research on the benefits of using edu-larp, for instance how it promotes empathy, identity exploration, problem solving, creativity and cooperation \cite{Maragliano2019Edu-larpGames,Lacanienta2022LiveLearning,caring,Mochocki2023Edu-LarpKnowledge},  

\subsection{What Larp Is and What It Is Not}
Larp is a versatile format, making it somewhat hard to pin down a clear definition \cite{Tychsen2006LiveSimilarities}. There have been several previous definitions formulated within the HCI community and in game design, for instance Salen and Zimmerman, already in their 2003 book \textit{Rules of Play - Game Design Fundamentals} \cite{Salen2003} provide a section about larp, where the definition of larp includes taking on personas of fictional characters, formal game statistics, narrative backstories, real physical spaces, acting out characters action, game mastering, social interaction and pursuing of narrative threads. Our definition in this paper mirrors contemporary trends in the larp community, and tries to encompass the complexity and variation of larps. There are an abundance of other definitions of larp and role playing games \cite{Harviainen2018Live-ActionGames, Zagal2018DefinitionsGames}. In this section we will formulate the definition of larp used in this paper.

In this paper, we refer to larp as co-creative improvisational roleplay, with a strong focus on role-taking, acted out embodied in the physical room, in a shared collaborative world building and story narrative, where there is no audience, but instead all participants take on roles and engage in the communal storytelling, thereby co-creating an experience. 

It should be noted right away that there exist many larps that do not fulfil all of these criteria – but together they frame not just the most common forms but also capture key aspects of larp that make them an interesting and unique art form. Role-taking is particularly important. Although physical aspects such as venue, costumes and props can play important roles for the immersion, larp is first and foremost a mental process, engaging with immersing into the story world and state of mind of a character, through an embodied first-person experience. Participants pretend to be someone else, with other feelings and agencies, in a different situation/fiction than the ordinary world around them.

Larping is typically done within some frames given by organizers, such as world setting and a collection of rules for the experience. Larps are thus designed, and have designers that may or may not also be players. Many kinds of larps exist, some are for a few persons such as the Stockholm Scenario Festival \footnote{Stockholm Scenario Festival, \url{https://scenariofestival.se/}} and others up to several thousand participants such as \textit{Mythodea} (see 15 in Appendix \ref{appendix: larps}). Larps can differ in duration from an hour up to several days or even weeks. Venues range from a simple classroom to a huge castle complex or an entire forest, and costumes range from simple t-shirts with prints on, to elaborate armor. There is also great variation in narratives, topics and tonality.

In this article we distinguish between leisure larps and edu-larps. Edu-larps in this context we define as when larp methodology is used for specific learning purposes \cite{Bowman2023EmployingCollaboration.}. Further, there are larp-like experiences, where the term larp is (commonly) not used, and often without clear connections to the larp community. Sometimes the word larp is even consciously avoided, for fear of the word being seen as not serious. Larp is also partly entwined with other similar sub-cultures such as megagames, cos-play, murder mysteries, some types of AR-games, pervasive games, reenactment, SCA (Society for Creative Anachronism) \footnote{Society for Creative Anachronism, \url{https://www.sca.org/}}, and escape rooms. Those can, especially from the outside, look a lot like larp. However, most often these larp-like experiences do not focus on role-taking, and therefore fall outside the scope of this article. This article will mostly focus on larps that are explicitly described as larp, but also to some extent engage with other larp-like experiences where the experience fits the definition, even if the term larp is not used.

There are also other kinds of activities in HCI that can be considered larp adjacent, such as designing for mixed reality performances \cite{Zhou2019Astaire:Players, Benford2011PerformingReality, Macaulay2006TheDesign, Nam2014InteractiveHCI, Taylor2017PerformingHCI} or embodied spectator experiences \cite{Tekin2017, Seering:2017:APG:3064663.3064732, Schnadelbach2008}. Simulations are also a genre that can be considered adjacent to larp---drawing on roleplaying for applied domains such as within medical fields or for disaster response teams \cite{ToupsDugas2016PlayingPractice, Bowman2014EducationalReview}. While these activities can be considered similar to larp in many ways, they differ primarily on the basis of the lack of the first-person perspective in most cases \cite{Striner2019ADomains}. Larp, different from performances, do not include an audience and participants have a high level of agency, as they are co-creating the story experience.

\subsubsection{Different types of larp designs}
 There is not one cohesive way a larp is designed, nor a single approach to writing larp scripts. Thematically, larps can vary greatly. The fantasy genre may dominate, but there are also larps that are set as sci-fi (see 4 in Appendix \ref{appendix: larps}), steampunk \footnote{Steampunk i Sverige, \url{https://steampunkisverige.wordpress.com/category/lajv/}}, post-apocalyptic (see Figure \ref{fig:blod}), Regency (see 9 in Appendix \ref{appendix: larps}), musical (see 13 in Appendix \ref{appendix: larps}), Wild West (see 20 in Appendix \ref{appendix: larps}), 1920s (see 18 in Appendix \ref{appendix: larps}), and many others. There are also different creative cross-overs, such as Jane Austen meets Westworld (see 19 in Appendix \ref{appendix: larps}). 

Just as in other forms of rpgs (roleplaying games), larps can be heavily scripted, often referred to as railroaded, or very free-form, often referred to as sandbox \cite{2014RailroadingSandbox, 2021SandboxBetter}. Larp holds game-like qualities, as there are a set of rules for each larp. The physical enactment of such rules help uphold the simulation of the game world. Rules are often also used to affect the narrative or the players' out-of-game knowledge. These types of rules are often referred to as meta-techniques \cite{Stark2014DefiningMeta-techniques}.

There are some main types of larp: \textbf{Gamistic battle larps} (american style larps) \cite{Unknown2022TheUS}, \textbf{Nordic larp}, a style which has grown to become an influential with a focus on communal storytelling and emotional experience over rules \cite{Saitta2014TheLarp}, and \textbf{Jubensha larp} which is a Chinese form of larp that has exploded in popularity during recent years and often played as murder mysteries with added roleplaying and immersive venues  \cite{xiong_player_2023, Xiong2022TheLarp}.

Larps, especially those in the Nordic larp style, can have transformative impacts on players, and are a strong method for engaging and creating emotions and transformation \cite{Bowman2022TransformativeCommunities}. Some larps are designed with the explicit purpose of (re)learning, practicing prosocial behaviors, and inspiring social change. For example, a non-profit larp company called \textit{Ursula} has made larps about enslavement, sustainable development, and feminism \footnote{F\"{o}renigen Ursula, \url{https://www.foreningenursula.se/}}, and \textit{Not Only Larp} is a company that makes immersive experiences for social change, such as larps focused on critiquing capitalism (see 23 in Appendix \ref{appendix: larps}).

\section{Method}
The methodology of this work, especially the synthesising section, can be characterized in the tradition of Humanistic HCI \cite{Bardzell2016HumanisticHCI}. Humanistic HCI, as with other well established humanistic research methods such as ethno-methodology \cite{OReilly2012EthnographicEdition}, auto-ethnography \cite{Marechal2010Autoethnography, Wall2006AnAutoethnography}, and narrative inquiry \cite{Riessman1993NarrativeAnalysis}, this approach embraces the subjectivity of studying human interactions and social phenomenons, thus not striving for objectivity. 

A group of experts engaged in the sensemaking and contextualization for this piece together, in an iterative process. During the span of several months, from October 2022 until August 2023, this expert group met around once a month and brought together knowledge on larp, and technology in relation to larp, through online meetings and conversations. At CHI'23 the group arranged the workshop called \textit{edu-larp@CHI}, focused on educational larps, bringing together 15 researchers with interest on the topic. This workshop, although not part of the data gathering for this paper, further spurred the discussions in the expert group. 

Over several months the expert group brought together and compared their knowledge on larp and larp related research, spanning several decades, and also presented and discussed larp research exemplars, as well as together identified research interests and topics relevant in relation to larp and HCI. Through these discussions the group identified knowledge on larp to synthesise, toward creating a foundation for future research in the HCI community. This included a literature review on larp-related work published within ACM, as well as a synthesis of knowledge from larp practice outside academia, and a set of annotated larp research exemplars, as presented in this paper. 

The paper consists of these three sections, that are then co-reflected upon in the discussion and synthesising section; a practitioner overview, a literature review and a number of larp research exemplars. Detailed methods for each of these sections are described in their respective section. 

The literature review conducted in section \ref{ACM} was limited to the ACM database, finding and thematically analysing larp-related publications. The practitioner overview builds on industry references mostly gathered through literature search as well as snowball sampling of industry references, based on the combined pre-knowledge in the expert group. The larp research exemplars are narrative-described first person perspective Research through Design exemplars, carried out by the different co-authors. They are chosen to try to represent some different examples on ways to do research combining larp and HCI, with the intent to be inspiration rather than an all-encompassing description of the larps themselves.

The final discussion and synthesising section of the paper brings together the three sections, identifies different relations between larp and HCI, identifies challenges and opportunities with using larp in an HCI research context, and points out possible directions forwards.   

\subsection{Positionality Statement}
The expert group consists of HCI researchers from the US and Europe, in different stages of their academic career, all who have published larp related research. Several of the authors are also active larpers and larp designers, with a long history of studying and arranging larps, within and outside the HCI context. At the time of writing, the expert group of researchers consisted of: 2 full professors (Sweden and the US), 1 associate professor (Sweden), (2 assistant professors (1 in Sweden, one in Spain), 1 postdoc (the US) and 2 PhD students (in the US and Sweden). All members of the expert group are co-authors of this paper. There is a limited amount of work done on larp in relation to HCI, and combined the expert group has been involved in a majority of the research in this field that has been done. 

The authors are, in most cases, insiders in different larp communities. This allows for deeper insights into larp, however we also acknowledge that this can create a bias toward our Western-centric perspectives of larp. Further, being the designers and larpers the authors have preferred methods, genres and larp styles, this especially affects the research design exemplars. The co-authors of this paper are located in North America and Europe, and have only directly experienced larp as it is created and played in these regions. While we don't have first hand experience with larp outside of NA and Europe, we know there is also robust Chinese and Brazilian larp communities, which we do make sure to include in our practitioner overview in Section \ref{PO}. The focus on Europe and the US limits the perspective of this synthesis paper to these traditions, apart from citing work that has been published in the HCI research literature. The authors' long term engagements with specific larp communities, as well as with larp related HCI research, has shaped their view on larp, and despite trying to provide an overview, we acknowledge that this overview is not objective or entirely disconnected from the positionalities of the authors. 

\section{Practitioner Overview of Larp, Technology, and Digital Aspects} \label{PO}
This section focuses on information gathered on HCI relevant aspects of larp outside formal academic publications. The information comes from online information sources, drawing upon prior knowledge within the expert group. This overview does not strive to be all encompassing or unbiased, as it is based on the group's knowledge. Still, within those limitations, the attempt is to represent larp as broadly as possible to give an understanding of at least some of what is out there. Larp in its essence can be seen as non technological, as it focuses on persons interacting in a physical space. But with the increased entanglement between the digital and physical, digital aspects have become more and more integrated into the larp community, larp organizing processes and the larps themselves. 

\subsubsection{Technology for arranging larps}
\begin{figure}[htbp!]
     \centering
     \subfigure{
         \includegraphics[width=0.4\textwidth]{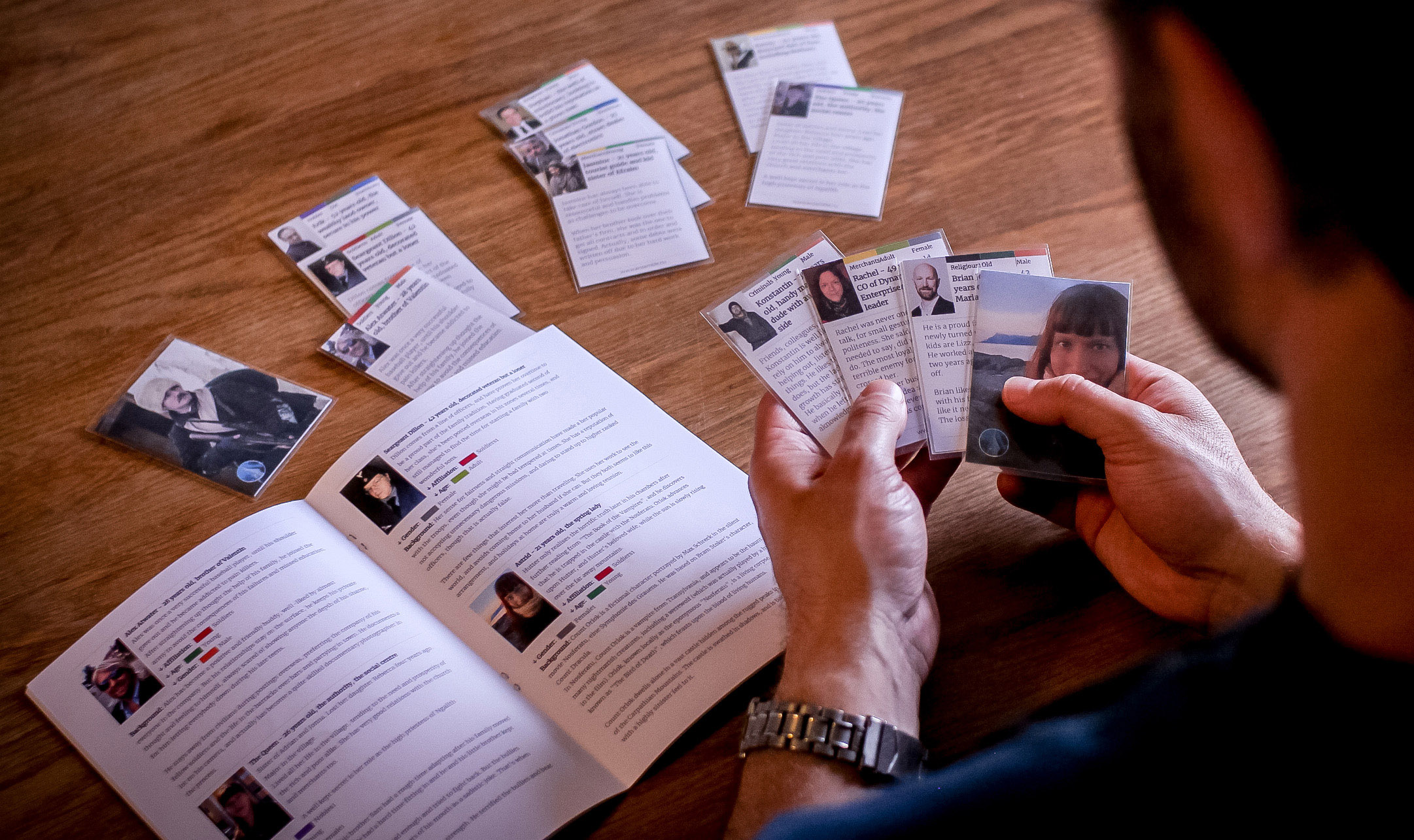}
         %\label{fig:ensemble}
         }
     \subfigure{
         \centering
         \includegraphics[width=0.4\textwidth]{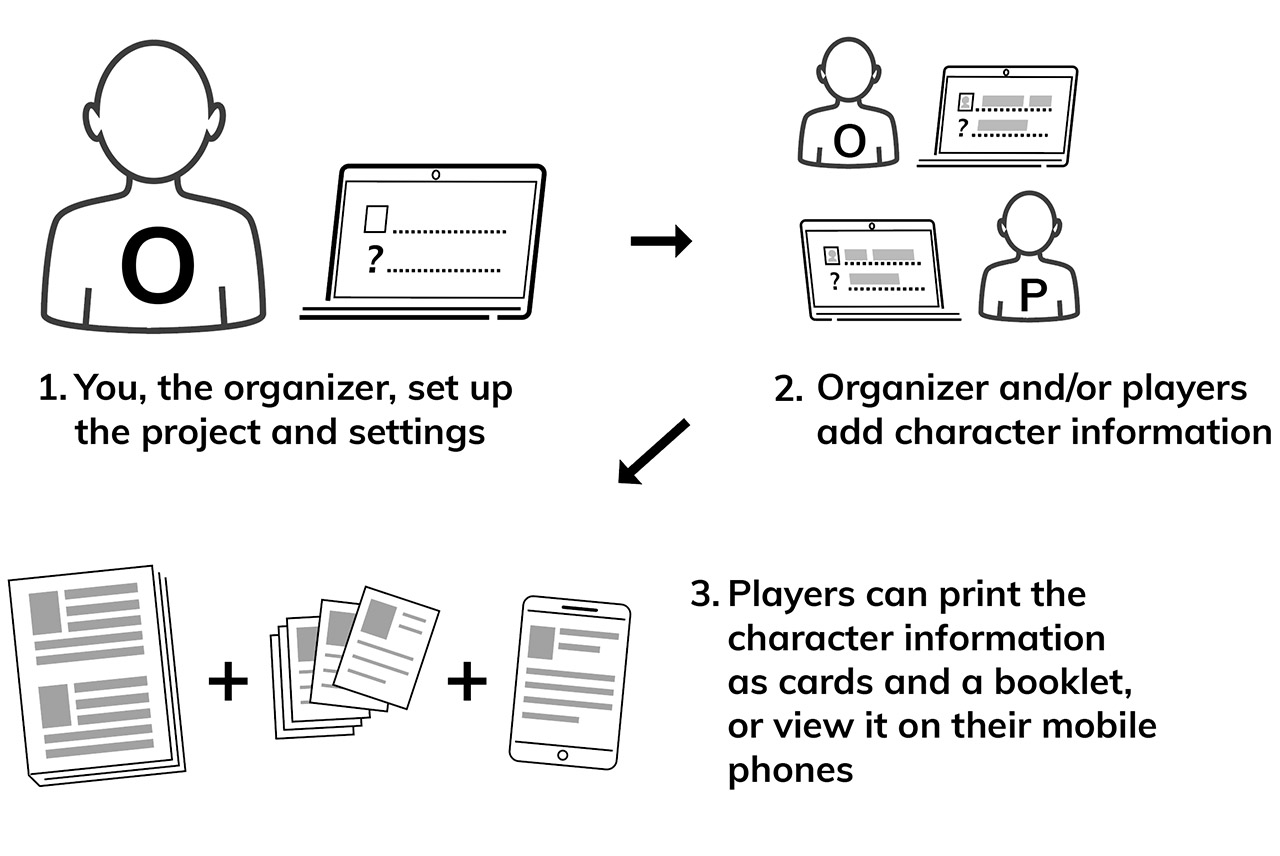}}
         %\label{fig:sketch}
        
        \caption[Ensemble]{Screenshots from the homepage for the digital character sheet tool \textit{Ensemble}, developed specifically to be used for larps. }
        \label{fig:ensemble}
        
\end{figure}

In the early days of larp, paper fanzines were often a way to spread information and exchange ideas. Since then, the larp community has been quick to adapt to new technologies used to facilitate both co-creation and community building, using email-lists, online forums, homepages, blogs, Facebook groups and Instagram accounts. Using IT-competences within the community, several unique digital tools for organising larps have also been developed \footnote{Underworld, \url{https://underworldlarp.com/managing-your-larp-technology/}}, for instance automatic scheduling systems for larp conventions \footnote{Prolog, a swedish larp convention \url{https://prolog2022.spelkonvent.se/}}, booking sites to buy larp tickets, and automatic systems to write, sort and distribute player characters \footnote{Ensemble, \url{https://ensemble.nu/}} (see Figure \ref{fig:ensemble}). For Jubensha larps, special apps are often used for signups and also to find and pair participants \cite{Xiong2022TheLarp}.

%\begin{figure}
%    \centering
%    \includegraphics[width=0.5\linewidth]{Images/ensemble.jpg}
%    \caption{Screenshots from the homepage for the digital character sheet tool ensemble, developed specifically to be used for larps}
%   \label{fig:ensemble}
%\end{figure}

%\begin{figure}
%    \centering
%    \includegraphics[width=0.5\linewidth]{Images/ensemble_sketch.jpg}
%    \caption{Screenshots from the homepage for the digital character sheet tool ensemble, developed specifically to be used for larps. The tool is for creating and distributing characters, automatically creating character cards as well as booklets, to give players an overview of characters at the larp}
 %   \label{fig:sketch}
%\end{figure}

\subsubsection{Technology in larps}
\begin{figure}[htbp!]
    \centering
    \includegraphics[width=0.5\linewidth]{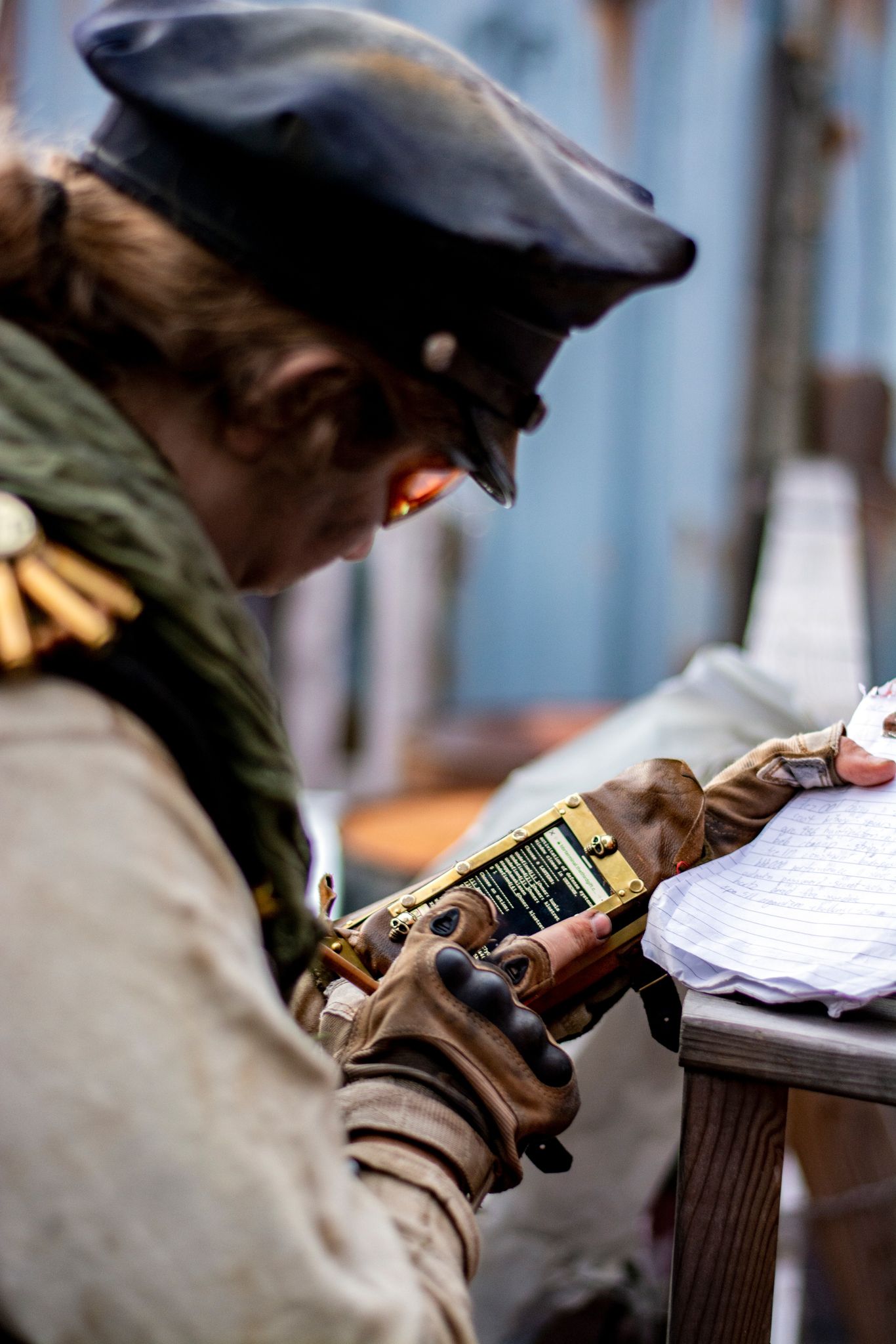}
    \caption[Blodsband]{\textit{Blodsband reloaded} larp. Here is a player using the in-game hacking program at the larp (see 7 in Appendix \ref{appendix: larps}). (Photo: Larpology)}
    \label{fig:blod}
\end{figure}

There have also been a lot of different examples of technology used in larps themselves. These can be game mechanics such as steering and interacting with other spaceships (see 22 in Appendix \ref{appendix: larps}), chat functions to for instance chat with NPCs (non-player characters) who are off-site (see 8 in Appendix \ref{appendix: larps}), tech used for simulating (for instance magic), tracking player locations, aesthetic reasons and more (see 14 in Appendix \ref{appendix: larps}). The tech can be diegetic (embedded into the larp narrative) or non-diegetic (in the background). Examples of technology and digital layers used are lights, soundscapes and smoke machines \cite{Segura2017DesignLarps}. There are also examples of robotics being used, for instance an animatronic Velociraptor running around in the bunkers of the survival horror larp \textit{Insomnia: Imperium} (see 12 in Appendix \ref{appendix: larps}. Jubensha larps utilise technology in many ways, for special effects and for instance to track player resources and stats \cite{Xiong2022TheLarp}. Some examples of technology in larps are shown in Figures \ref{fig:blod} and \ref{fig:horizon}. Utilizing technologies at larps turns them into a type of hybrid experience, that changes the way we should think about designing the experience. Hybrid experiences, which mix physical, digital and technological components requires special design considerations, as has been seen in for instance research on hybrid games \cite{Sparrow2023LessonsPlay}.

There are organisations focusing specifically on developing technologies to be utilized at larps, such as Frivolous Engineering \footnote{Frivolous Engineering, \url{https://www.facebook.com/FrivolousEngineering/}}, a non-profit organization based out of the Netherlands that designs technology for larps \cite{Segura2017DesignLarps}. The technologies developed for larps can be used to create effects for immersion such as sound, lights and smoke, but they are also utilized to drive plots forward, contribute to character development, changing character agencies and steering the narrative. 

\begin{figure}[htbp!]
    \centering
    \subfigure[Photo by Kai Simon Fredriksen]{
    \includegraphics[width=0.5\linewidth]{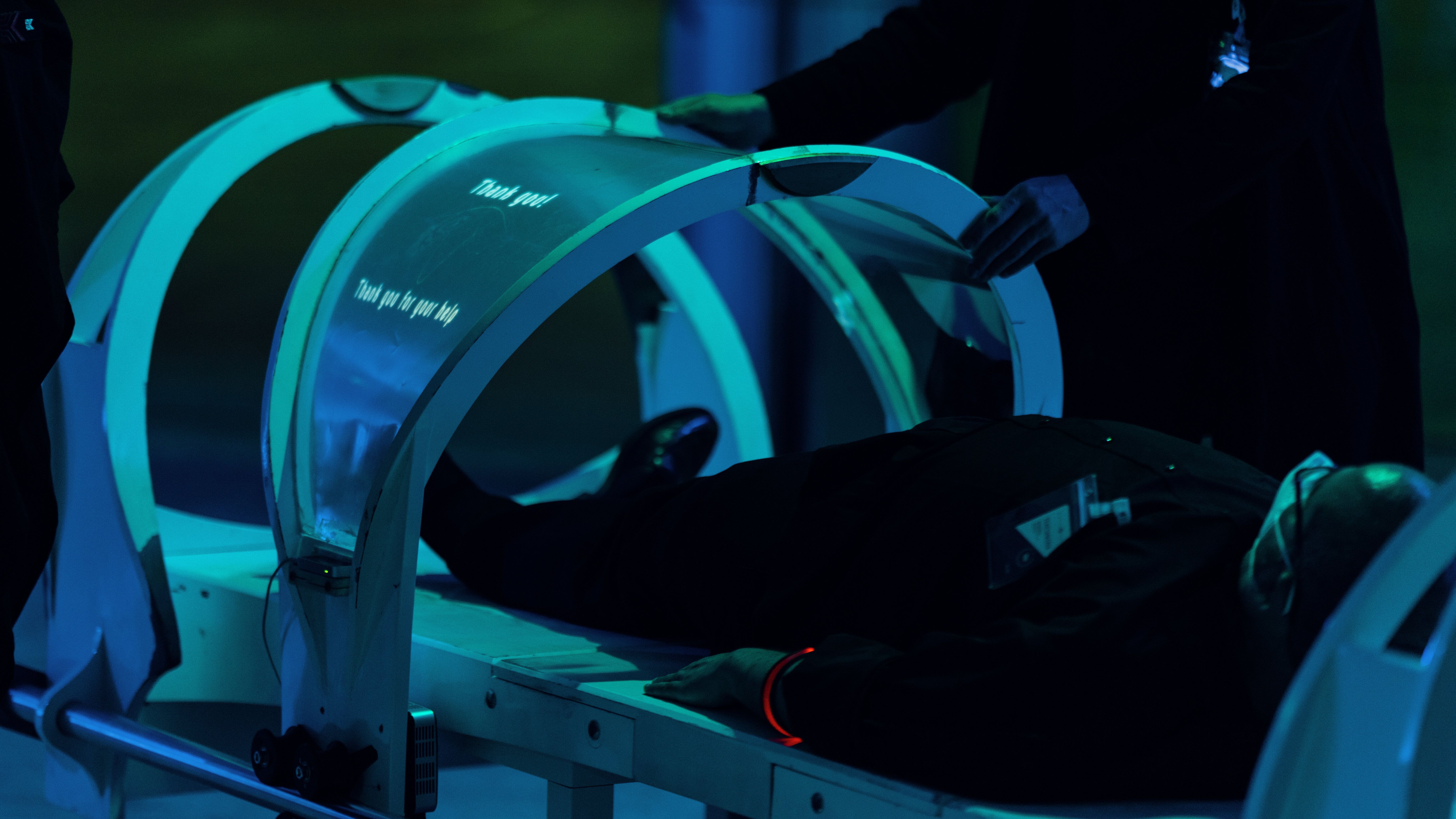}
    \label{fig:cmachine}
    }
    \subfigure[Photo by Sanna Vuorenmaa]{
    \centering
    \includegraphics[width=0.5\linewidth]{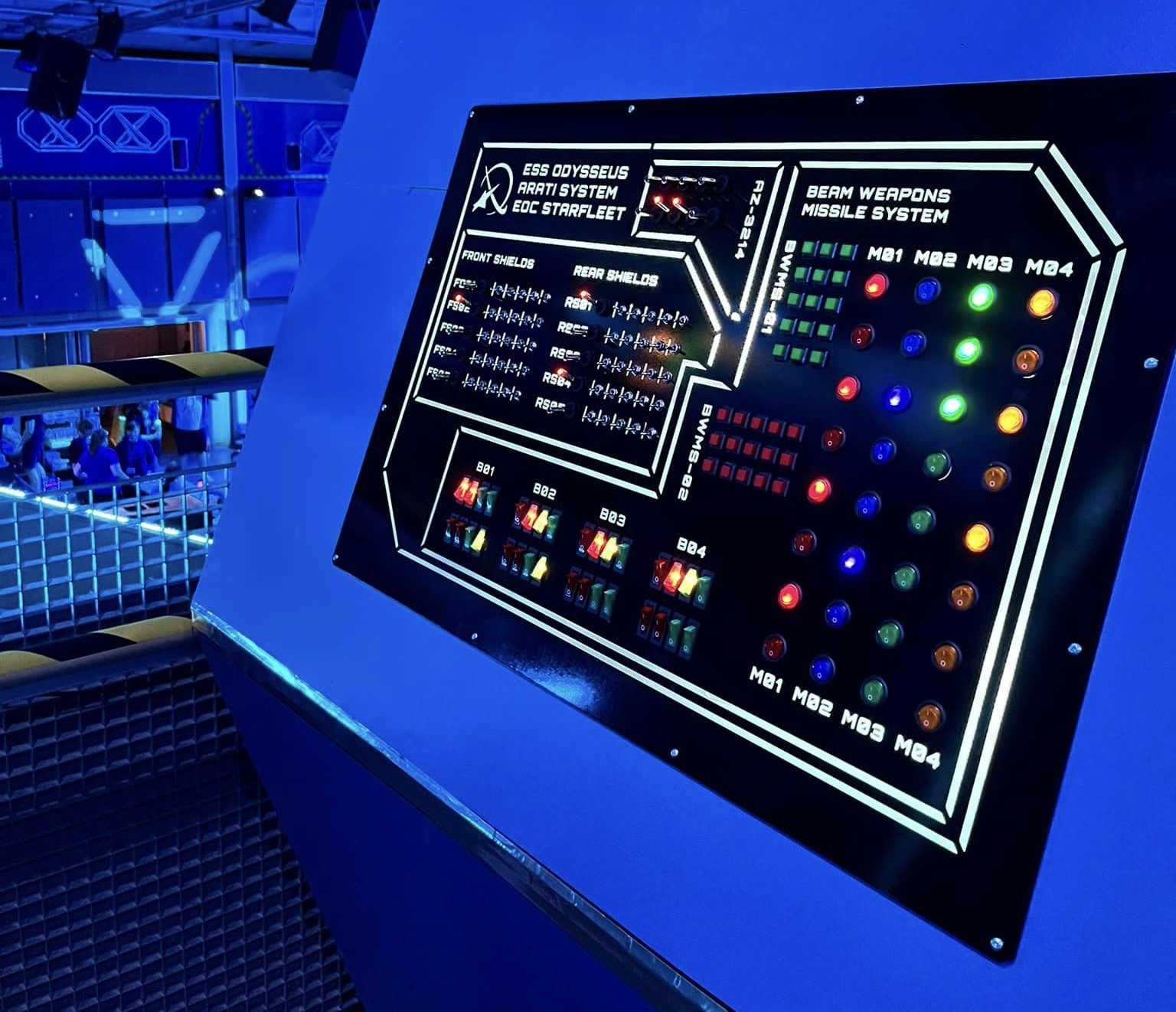}
    \label{fig:vissers}
    }
    \caption[Frivolous Engineering Tech]{Technology designed for and used at various larps. Image (a) is of a medical facility used at the futuristic larp called Mission Together (see 26 in Appendix \ref{appendix: larps}) and Image (b) is of a control station from the larp Odysseus (see 25 in Appendix \ref{appendix: larps}).}
    %\label{fig:marie}
\end{figure}

A common solution at larps is that the technology is operated by Non-Player Characters (NPC), such as phones where you call a characters relatives, voiced by an NPC. Or, alternatively, larps use more advanced props such as faked medical facilities (see Figure \ref{fig:cmachine} from the larp \textit{Mission Together}), a machine for faking consciousness transfer at the larp Greylight 2142 (see 10 in Appendix \ref{appendix: larps}), and the control stations at the larp \textit{Odysseus} (see Figure \ref{fig:vissers} and further detailed in entry 26 of Appendix \ref{appendix: larps}). But there are also technologies that are self-sustained and can be operated directly by the players, for example the fortune telling device made by the larp prop company \textit{Frivolous Engineering}, where characters received ominous messages about their future, used for instance at the larp \textit{Cirque Noir} \footnote{Cirque Noir \url{https://nordiclarp.org/wiki/Cirque_Noir}}.

\subsubsection{Fake technology in larps}
\begin{figure}[htbp!]
    \centering
    \includegraphics[width=0.425\linewidth]{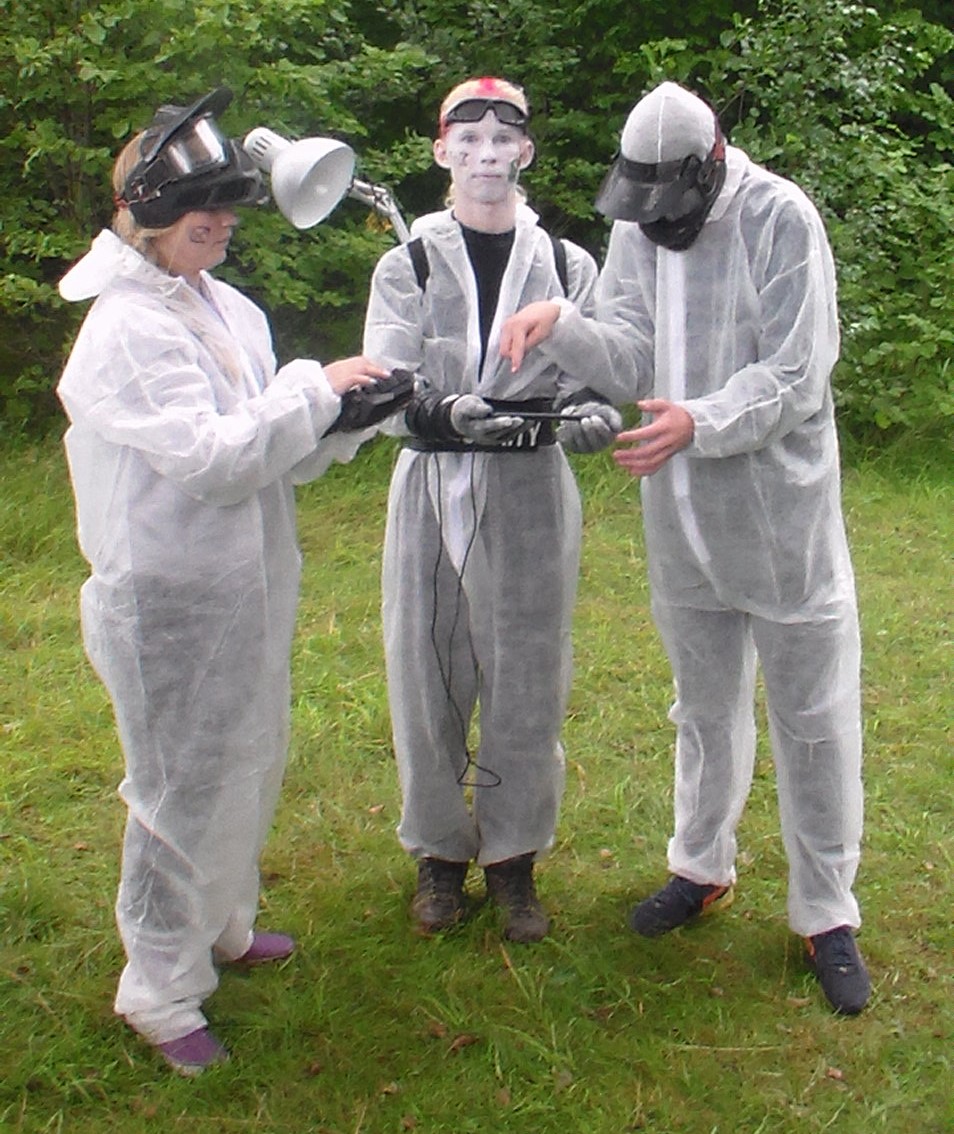}
    \caption[Shows \textit{Mutanternas tidsålder} larp, (Photo: LajvVerkstaden)]{Shows \textit{Mutanternas tidsålder} larp, (Photo: LajvVerkstaden), see 14 in Appendix \ref{appendix: larps}}
    \label{fig:mutant2}
    %\label{fig:marie}
\end{figure}
Another way that larp incorporates technology is by creating and using fake technology. Larp's immersive qualities and co-created storytelling make it possible to co-build a fiction that includes trying out technology that does not exist, and seeing how it affects persons, groups, lifestyles and society. This kind of fake technologies and simulations bear resemblance to design methods such as Wizard of Oz \cite{Kelley2018WizardJourney} and rapid prototyping. A common example of fake technology at larps is people playing androids (example in Figure \ref{fig:mutant2}), or pretending to have been bio-hacked (see 3 in Appendix \ref{appendix: larps}). An example of this was the larp \textit{Last Will} (see 24 in Appendix \ref{appendix: larps}), a larp focusing on modern and future slavery. The larp was set in a dystopic future, where people in debt could become ``lifers'' - that is slaves. Slaves got a necklace that the owners could remotely activate to deliver an electric shock, or just send out warnings of a shock. Shocks could be dealt individually, but also to the whole group of lifers, resulting in a collective punishment. This became a powerful game mechanic, despite it in reality just being simple led-light necklaces, and the ``remote controls'' being ordinary whistles. It worked because everyone accepted and play-acted the fiction.

%\begin{figure}[htbp!]
 %   \centering
  %  \includegraphics[width=0.5\linewidth]{Images/android_crop.jpg}
   % \caption[Mutanternas tidsålder]{\textit{Mutanternas tidsålder} was a several weeks long summer camp larp where the participants play mutants and androids in a world controlled by a gigantic underground computer. In this picture is an NPC is playing an android, that participants programmed by touch and voice commandos\footnote{\url{https://lajvverkstaden.se/}} (Photo: LajvVerkstaden).} 
%    \label{fig:mutant}
%\end{figure}

%\begin{figure}
%    \centering
 %   \includegraphics[width=0.5\linewidth]{Images/Fight Like a Girl larp. cred Jan-Åke Fonnaland.jpg}
  %  \caption[Bio-hacked soldiers]{Bio-hacked soldiers, the biohacking being represented by simple means such as face paint, glasses and gadgets, in the feminist futuristic larp %\textit{Fight Like A Girl} \footnote{Fight Like a Girl, %\url{https://flaglarp.wordpress.com/}} (Photo: Jan-Åke Fonnaland)}
%    \label{fig:soldiersfightgirl}
%\end{figure}

\subsubsection{Larps about technology}
There are many examples of larps exploring technology enhanced realities (see 6 in Appendix \ref{appendix: larps}), such as dystopian, utopian, and what-if worlds, as in sci-fi, steampunk and post-apocalyptic genres.

\begin{figure}[htbp!]
    \centering
    \subfigure{
    \includegraphics[width=0.4\linewidth]{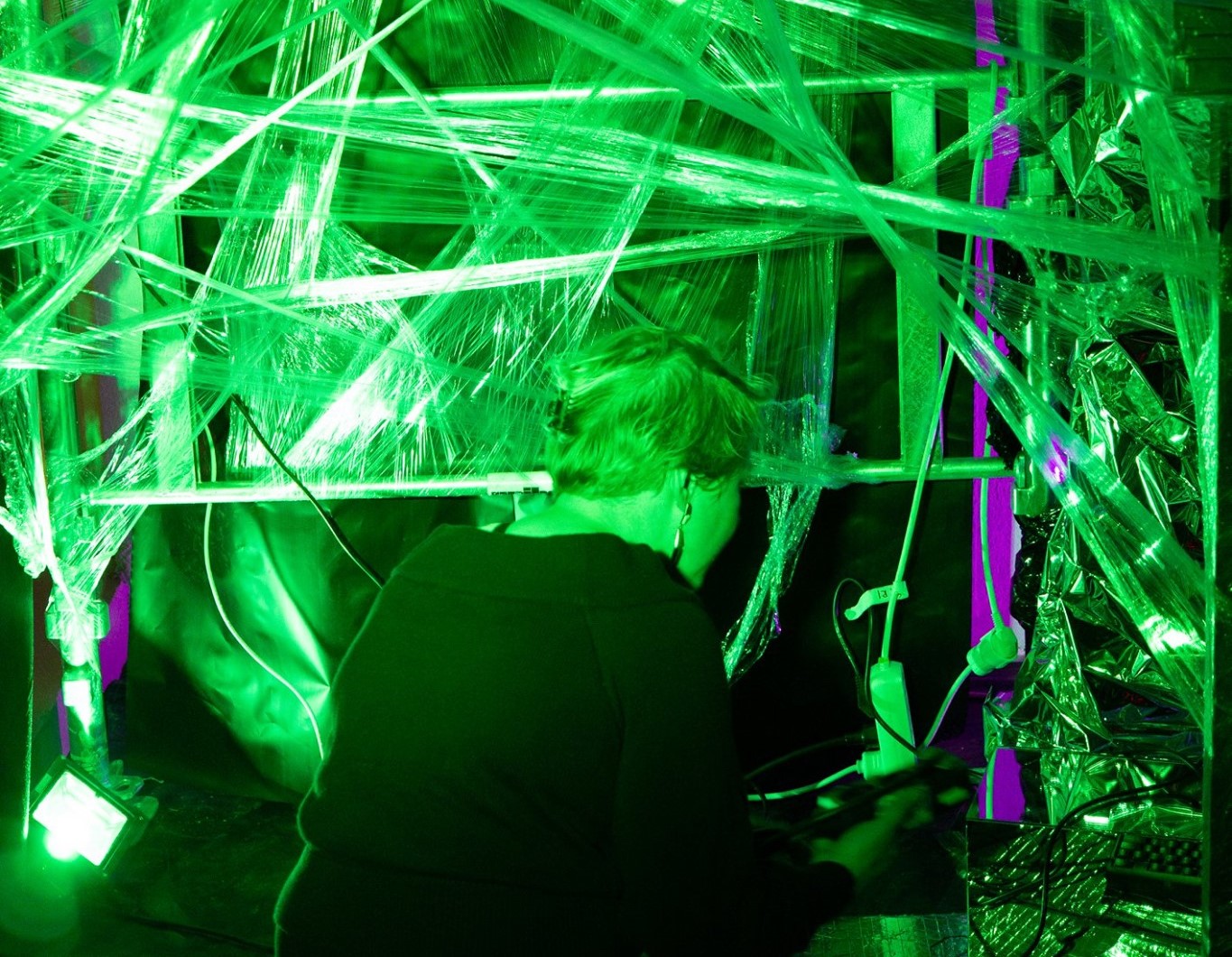}
    }
    \subfigure{
    \centering
    \includegraphics[width=0.465\linewidth]{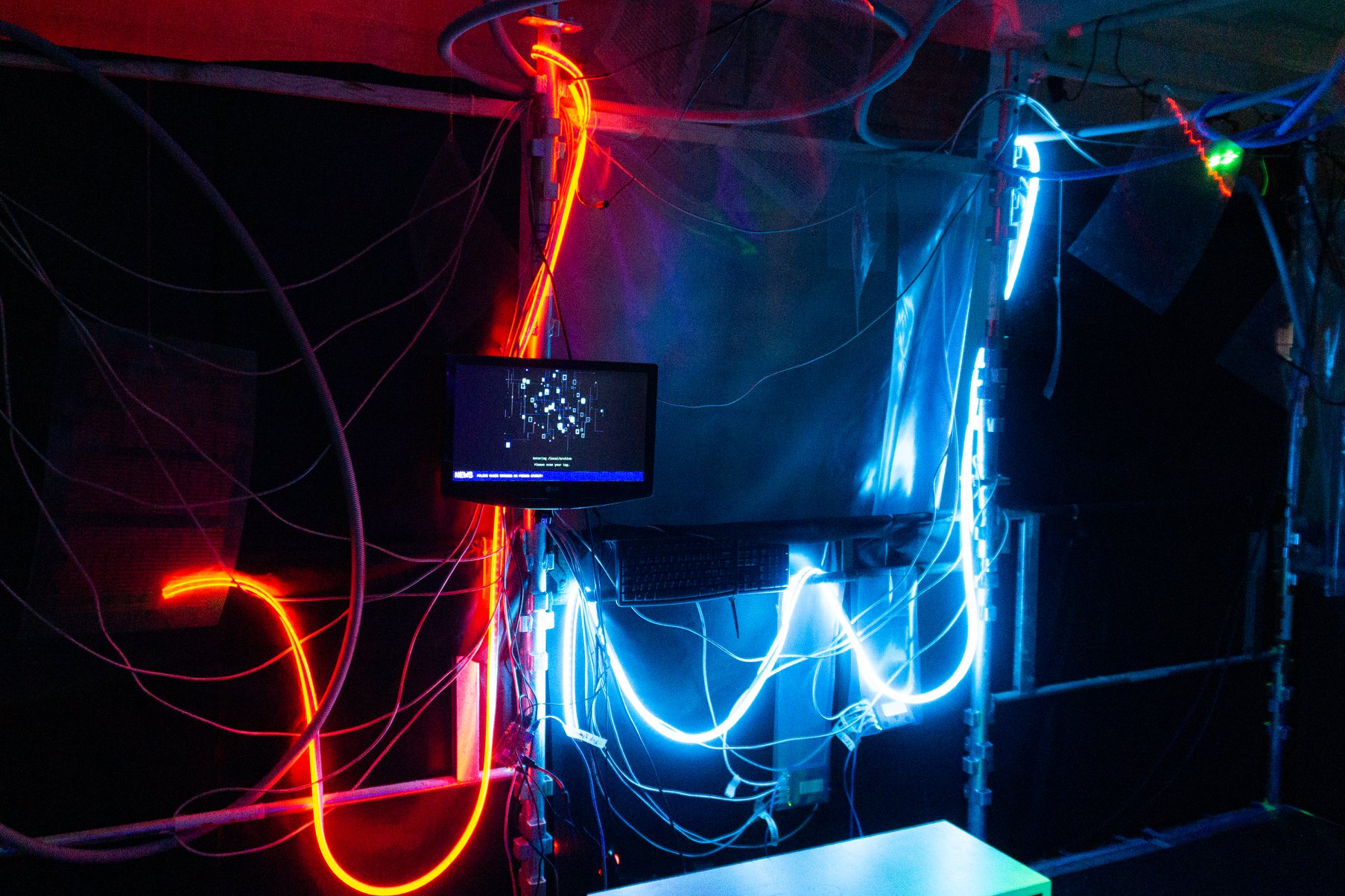}
    }
    \caption[Beyond the Neural Horizon]{In \textit{Beyond the Neural Horizon} everyone played hackers - or rather their digital avatars, but in a physical room. The venue was a physical manifestation of the internet.  You played a digital version of yourself, treating digital technology as a world in itself (see 6 in Appendix \ref{appendix: larps}). (Photo: Atropos)}
    \label{fig:horizon}
\end{figure}

\begin{figure}[htbp!]
    \centering
    \includegraphics[width=0.5\linewidth]{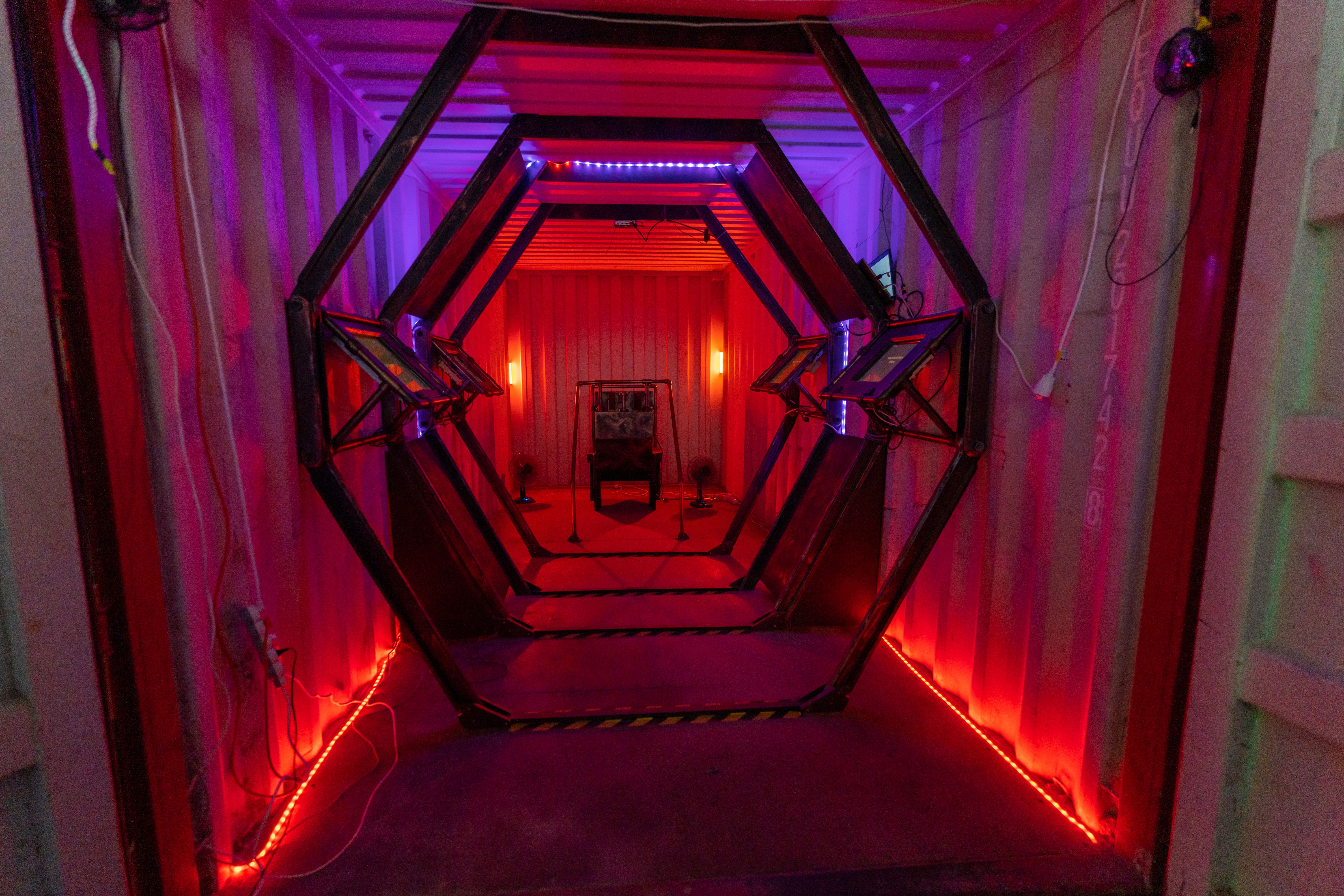}
    \caption[MT]{Venue for the larp \textit{Mission Together} - a sci-fi larp exploring future societies, arranged by Not Only Larp (Photo: Kai Simon Fredriksen). See 26 in Appendix \ref{appendix: larps} for more information about the larp.}
    \label{fig:Zero}
\end{figure}

\begin{figure}[htbp!]
    \centering
    \includegraphics[width=0.5\linewidth]{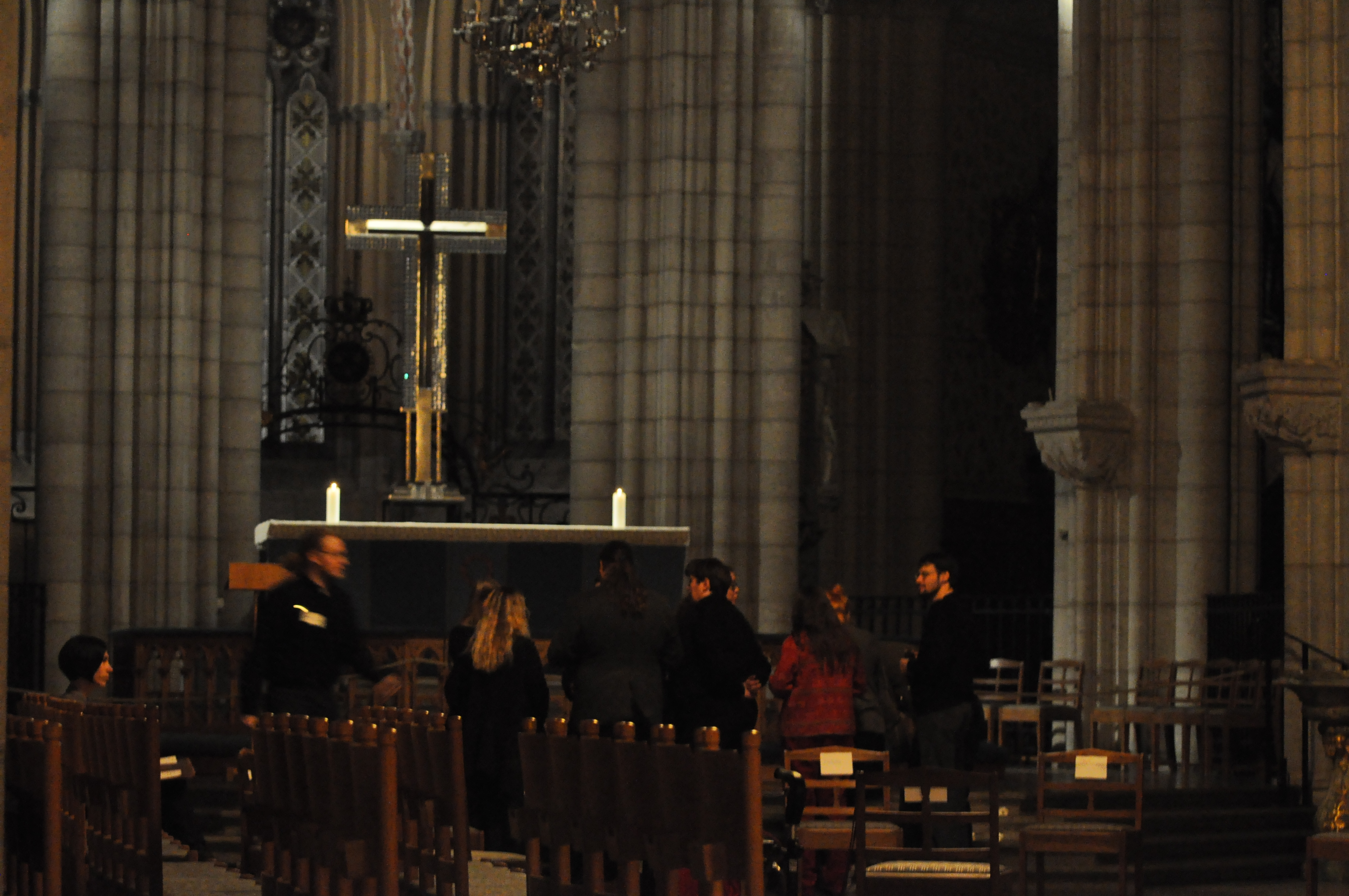}
    \caption[Turings Fråga]{\textit{Turings Fråga} (see 21 in Appendix \ref{appendix: larps}) was run in the Arch-Cathedral of Sweden, in collaboration with the Swedish Church, organized by Dahlberg, Engström and Gamero. The larp was set in a non-defined future, and the participants were trying to pass a test to enter a prestigious academy. The test consisted of trying to identify the android in the group, and this was done using different criteria of what it means to be human (Photo: Johan Dahlberg)}
    \label{fig:fraga}
\end{figure}

\begin{figure}[htbp!]
    \centering
    \includegraphics[width=0.5\linewidth]{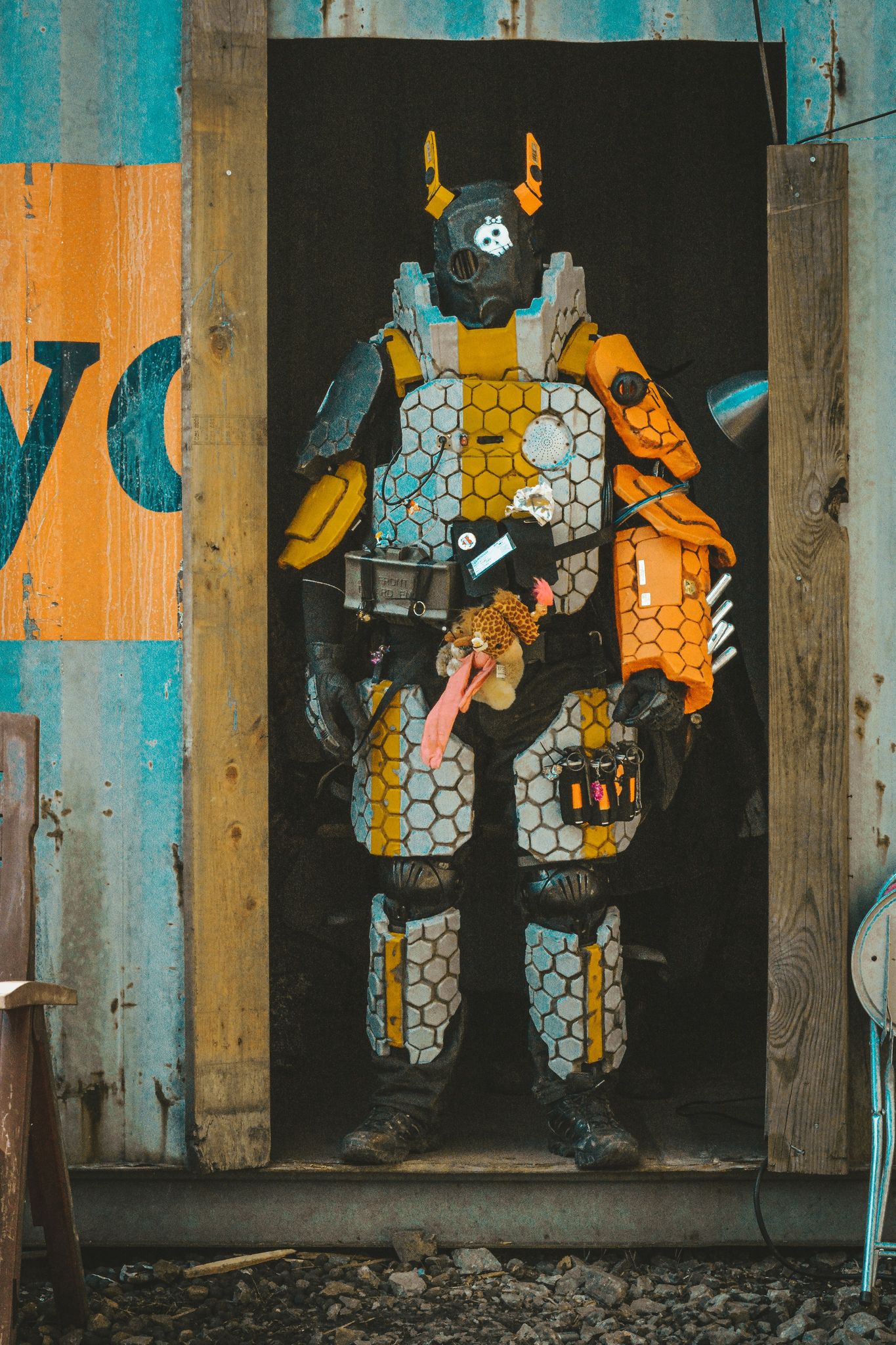}
    \caption{Participant playing a robot at \textit{Blodsband Reloaded} (Photo: Larpology)}
    \label{fig:blodsband}
\end{figure}

Some larps have technological exploration as a main focus. For instance \textit{Omgiven av idioter} (see 17 in Appendix \ref{appendix: larps}, translated to `Surrounded by idiots', in which participants lived in a near future where they lived all their lives via social media, and therefore needed special training in meeting people in real life (IRL). In this larp they were given devices that could, according to the fiction, show their emotions, color coded, as a wearable. Many larps engage with portraying and exploring different technological futures, and some larps dwell especially on philosophical and existential implications of technology. Examples of larps about technology are shown in Figures \ref{fig:horizon}, \ref{fig:Zero}, \ref{fig:fraga}, and \ref{fig:blodsband}.

%begin{figure}
 %   \centering
  %  \includegraphics[width=0.25\linewidth]{Images/android character - cred susanne vejdemo.jpg}
   % \caption{Enter Caption}
   % \label{fig:enter-label}
%\end{figure}
\subsubsection{Larps through technology}
Despite the fact that larps are generally focused on acting out stories together in a physical room, there have been many examples of larps playing with the format and creating different hybrid concepts using technology. This was further propelled by the Covid-19 pandemic, during which physical larps were largely discontinued. 

This type of experimentation with the format, and creating digitally hybrid larps, has been done using a wide range of different technological solutions. For instance there have been larps over discord \cite{how_2021}, Zoom-larps \cite{how_2021}, larps in VR (see 2 in Appendix \ref{appendix: larps}) and larps in Second Life \footnote{Playable Theatre, \url{https://www.indiecade.com/slarp-fest-2021/}}. An online specific genre has even sprung up: online larp, or LAOG (Live Action Online Games) \cite{Reininghaus2021ThreeLAOGs}, and there have also been academic courses on how to run online larps \footnote{ LajvVerkstaden, \url{https://lajvverkstaden.se/online-kurs-for-lajv-online-sodertorns-hogskola/}}. During Covid-19 Jubensha apps were launched and became very popular in China \cite{Xiong2022TheLarp}.

While some of these concepts push the limits of what can be seen as larp, they clearly stem from the larp design tradition, and with roleplaying and social co-creative storytelling as a core. In some cases, for instance with larps over Zoom, the use of technology can be diegetic, meaning you dress up as and play a character that is sitting in front of a screen, so that this is a shared situation for both player and character \cite{how_2021}. An example of larps through technology is shown in Figure \ref{fig:alchemist}, as well as in the larp \textit{Ancient Hours}, detailed further in entry 2 of Appendix \ref{appendix: larps}.

\begin{figure}[htbp!]
    \centering
   \includegraphics[width=0.46\linewidth]{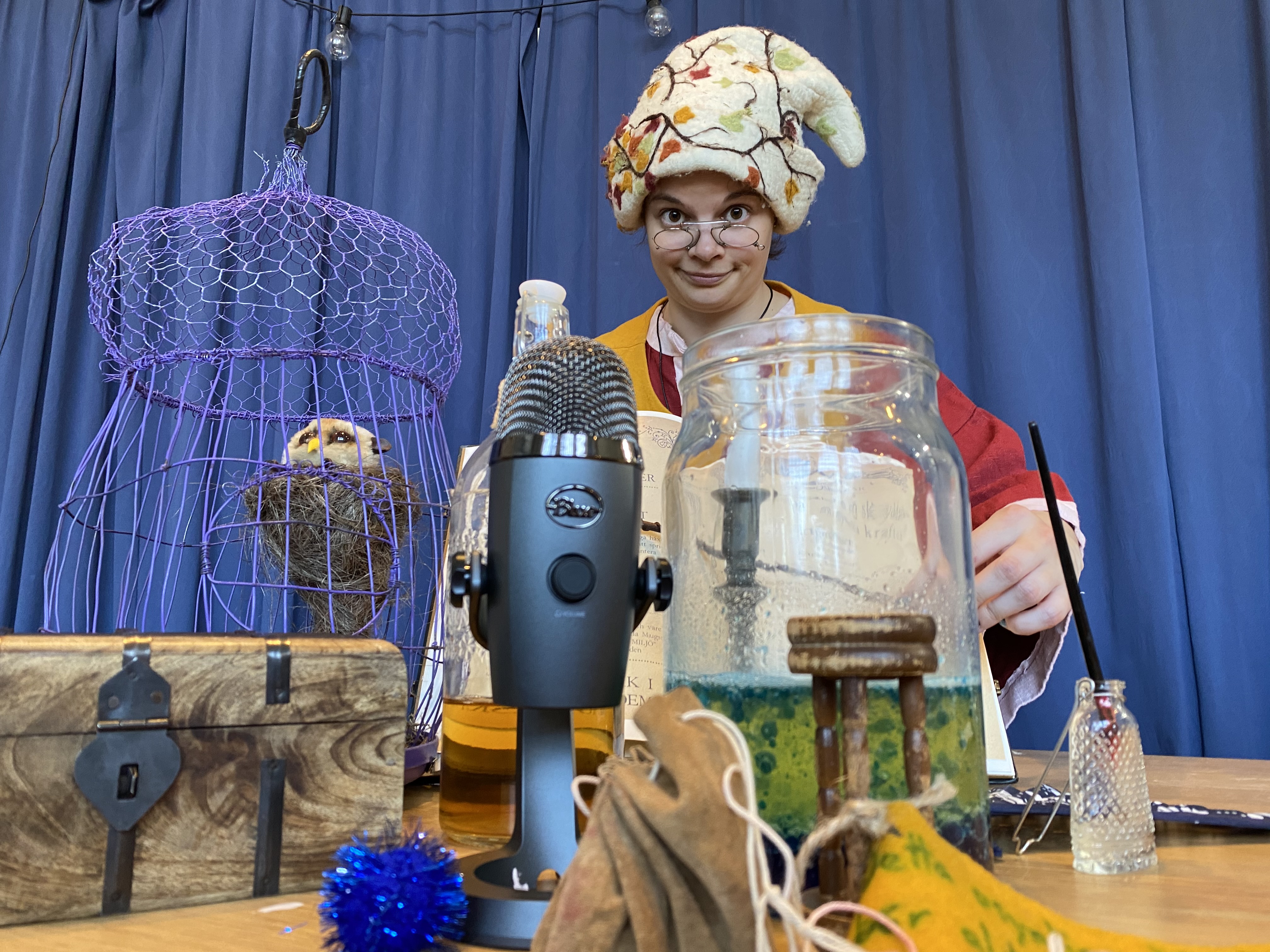}
    \caption[two images]{The alchemists classroom (note the microphone in the front). (Photo by LajvVerkstaden).}
     \label{fig:alchemist}
\end{figure}

%\begin{figure}[htbp!]
%    \centering
%    \includegraphics[width=0.5\linewidth]{Images/IMG_0944.jpg}
%    \caption{The alchemists classroom in an educational online larp, note the microphone in the front. (Photo: LajvVerkstaden)}
%    \label{fig:alechemist}
%\end{figure}

%\begin{figure}[htbp!]
%    \centering
%    \includegraphics[width=0.5\linewidth]{Images/JoNadjaAHbyMilesLizak_crop.jpg}
%    \caption[VR]{The VR-larp Ancient hours (photo: Miles Lizak), \url{https://alexandria.dk/en/data?scenarie=14058}}
%    \label{fig:ancient}
%\end{figure}

\section{Literature Review: Larp in ACM} \label{ACM}
As mentioned, there are many larp-adjacent concepts in our field, but considering larp as a cultural phenomena and specific method, it is of interest to look at what has been done explicitly exploring larp in relation to HCI research. Thus, as part of the research process for this paper, a literature review was conducted in June 2023 using the ACM Digital Library. The inclusion criterion was that the term ``larp'' was used anywhere in the paper (body, title, keywords, etc.). ``Larp'' was used as the keyword in the search. All types of publications were included, and all years. For the purpose of the paper, it was important to narrow the scope of roleplaying to the cultural phenomenon of larp. Attempts were made to search also for related terms, such as roleplay, which have been used in many forms within the design and the games communities, but due to the wide use of the words roleplay, it was hard to untangle larp and larp-like experiences from non-larp, and therefore the inclusion criteria of explicitly using the word larp was settled upon. This first step resulted in 123 papers. 

In a next step, all the articles were examined toward excluding irrelevant matches. The exclusion criterion (eliminating papers in which the abbreviation larp did not refer to live action roleplay), resulting in 71 papers remaining. After the selection process the abstracts and keywords for all the articles were read through, identifying HCI relevant concepts and terminologies mentioned. Themes were identified through a combination of keywords found in the articles, as well as based on the authors' previous knowledge on topics related to larp, thus it is a combination of bottom-up and top-down thematic analysis.

The purpose of this literature review was to understand what topics and themes within HCI surrounding larp are frequently studied, in order to provide an overview of the field and identify future research directions and opportunities. Our review contains elements of both a systematic literature review \cite{Moher2009PreferredStatement} as well as a systematic mapping review \cite{Epstein2020MappingLiterature, Grant2009AMethodologies}.

\subsection{Amounts and Publication Years}
The first ACM article briefly mentioning larp is from 2000 \cite{Crawford2000Forum}. Next, and the first actually engaging with larp, is  from 2004 \cite{Magerkurth2004AugmentingTechnology}. After this, we see a slowly growing stream of publications about larp in our research community. There is an increase in articles mentioning larp in later years, going up from around 2-3 per year to around 5-10 per year by 2017. Some articles have the same or overlapping authors, with around 20 authors in total, from around 20 different affiliations/universities, that write about larp. The vast majority of articles are centered on European and North American larps. 

\subsection{Central Themes and Topics in the Publications}
All mentions of HCI key concepts and terminologies in abstracts and keywords were listed, resulting in a list of major topics that over the years have been the focus of articles mentioning larp. A list of the publications can be found in the supplementary material of the paper. 
\begin{itemize}
    \item Pervasive games/play \cite{Oliveira2013EctodiegesisGames, Jegers2009ElaboratingEnjoyment, Calvillo-Gamez2011EmpiricalModel, Irie2017PervasiveLoci, Ahn2016PervasivePlay, Jonsson2008TheGames, Jonsson2006Prosopopeia:Larp, Waern2009TheGames}
    \item Playful interaction \cite{Pothong2021DeliberatingActionRolePlay, Pichlmair2017DesigningPlay, AltarribaBertran2019MakingInteraction, Spiel2019ThePlay, Waern2009TheGames, Buruk2019AWearables, AltarribaBertran2020TechnologyAgenda}
    \item Embodied interaction \cite{Vega-Cebrian2023DesignCharacterization, Beuthel2022ExploringPractices, MarquezSegura2019LarpingMethod, Dubbelman2011DesigningGames, Zhou2019Astaire:Players, Isbister2015}
    \item Wearables \cite{Buruk2019AWearables, Buruk2021TowardsWearables, Isbister2015, Dagan2019DesigningVulnerability, Dagan2019AVulnerability, Dagan2020FlippoCompanion, Alharthi2018DesigningPractice, MarquezSegura2018DesigningDesigners, Jung2021DesigningCreation, Dagan2019WorkshopVulnerability}
    \item Design fiction/futuring \cite{Ferri2019TakeFiction, Hutchison2007BackReality, MarquezSegura2018DesigningDesigners}
    \item Augmented reality \cite{Wetzel2008GuidelinesGames, Buruk2016WEARPG:Devices, Calvillo-Gamez2011EmpiricalModel, Magerkurth2004AugmentingTechnology}
    \item Appropriation \cite{Segura2017DesignLarps}
    \item Serious games \cite{Becker2007OhDesign, Medler2010TheMethodologies}
    \item Situated action \cite{AltarribaBertran2019DesigningDesign, AltarribaBertran2020TechnologyAgenda}
    \item Design methodologies \cite{Medler2010TheMethodologies, Kleinen2018UsingEnvironment, Jung2021DesigningCreation}
    \item Communication \cite{Drachen2008PlayerGames, Alharthi2018DesigningPractice, Becker2007OhDesign, Ciancia2013TransmediaProduction}
    \item Virtual Reality \cite{Zhou2019Astaire:Players, Lobser2017FLOCK:Experience, Hutchison2007BackReality}
    \item Value in design \cite{Jerrett2022ValuesSpace, Pothong2021ProblematisingDeliberation, Calvillo-Gamez2011EmpiricalModel, MarquezSegura2018DesigningDesigners, AltarribaBertran2020TechnologyAgenda}
    \item Immersion \cite{vejdemo-johansson_demo_2014, Spiel2019ThePlay, Calvillo-Gamez2011EmpiricalModel, Waern2009TheGames}
    \item More than human \cite{Spors2023LongingTechnologies, Light2023InOthers, Biggs2021WatchingDesign}
    \item Human centered design \cite{Pothong2021ProblematisingDeliberation}
    \item Interactive storytelling \cite{Tychsen2006RolePlatforms, Dubbelman2011DesigningGames, Harley2022TogetherNarrative}
    \item Tangibility \cite{Harley2022TogetherNarrative, Buruk2017AugmentedDevices}
    \item Participatory design \cite{Ciancia2013TransmediaProduction, Jung2021DesigningCreation}
\end{itemize}

\subsection{Common and Less Common Topics}
Larp as a tool in design processes seems to be the most common type of paper about larp in our community. Interesting to note is that leisure larps are written about more often in the earlier years, but almost nothing is written about them in later years \cite{Jonsson2006Prosopopeia:Larp, Dagan2019AVulnerability, Wohn2017ASports, vejdemo-johansson_demo_2014, Segura2017DesignLarps, Pichlmair2017DesigningPlay, Dagan2019DesigningVulnerability, Oliveira2013EctodiegesisGames, MarquezSegura2018DesigningDesigners}. It is also worth noting that not a single paper, as found through this literature review, seems to focus on larps for formal education, despite this being a common type of edu-larp. There is a heavy slant towards focusing on adults and professionals. There is one article that focuses on children at a summer larp camp for girls \cite{Fey2022AnywearGirls}. 

Concepts such as pervasive games/pervasive play are more common in the earlier articles, while topics like wearables, design fiction and embodied interaction have been more common in later years. The research on larp seems to pick up on trends in the HCI community, utilising larps for different purposes and to explore current central topics. Most articles focus on designing larps, with the explicit aim of doing research. There seems to be a lack of articles looking at existing leisure larps, for instance doing interviews and observations of larpers and larp designers. One notable exception is the FDG'17 paper \textit{Design, Appropriation, and Use of Technology in Larps} \cite{Segura2017DesignLarps}, consisting of a survey of 39 larpers, investigating how they use digital technology in larps, finding around 300 examples of technology used, for different purposes and of different kinds. 

\subsection{Relating Publications to HCI Theories}
It is noteworthy that publications focusing on larp seldom try to disentangle concepts and theories in relation to one another. Instead, most include a short section of background literature, and then move on to more practical design research. In particular, there is not much outward looking when it comes to the epistemological and theoretical foundations of this larp research, rather the references used often stay within the field, and mostly refer to more practically useful concepts bordering or directly tied to the larp research, typically referring mostly to more contemporary research. 

\subsubsection{Research through Design} This is an approach used in many of the publications, tying into works done by for instance Frayling \cite{Frayling1993ResearchDesign} and Zimmerman et al. \cite{Zimmerman2007ResearchHCI}. In practice, this means that larps are designed as part of the research process, thus creating knowledge. Those papers often seems to produce some type of design implications, creating strong concepts and intermediate level knowledge as formulated by Löwgren and Höök \cite{Lowgren1999DesignPractice}, or relating to ultimate particulars, as formulated by Stolterman \cite{Stolterman2008TheResearch}. One paper (\textit{Activity as the Ultimate Particular of Interaction Design} \cite{Waern2017ActivityDesign}) notes that in larps there is an activity design focus, rather than a focus on artifacts.

\subsubsection{Futuring} This common theme ties into the research traditions of design fiction \cite{Dunne2013SpeculativeDreaming} and critical design \cite{Bardzell2018CriticalBlythe.}. For instance, exploring the possibilities of wearables in relation to future tech and vulnerability \cite{Dagan2019DesignWearables}. 

\subsubsection{Value-Based Design Research} Research combining larp and societal issues has tied into variations of value-based design research \cite{Friedman2019ValueDesign}. An example is the DIS'21 publication \textit{Problematising Transparency Through LARP And Deliberation} \cite{Pothong2021ProblematisingDeliberation}.

\subsubsection{Embodied Interaction, }as formulated by  for instance Dourish \cite{Dourish2001WhereInteraction}, plays a role in several publications, such as the DIS'19 publication \textit{Larping (Live Action Role Playing) as an Embodied Design Research Method} \cite{MarquezSegura2019LarpingMethod}. Social embodied interaction has been a particular focus, for example in the CHI EA'18 paper \textit{Firefly: A Social Wearable to Support Physical Connection of Larpers} \cite{Vanhee2018Firefly:Larpers}. 

\subsubsection{Playfulness} This is a recurring theme in many publications. Play has been a focus in HCI communities for a long time, even resulting in the creation of a specific venue for play-focused researchers, CHI PLAY \footnote{CHI PLAY 2023 website, \url{https://chiplay.acm.org/2023/}}. Regarding stances and theories within HCI, Ludic design \cite{Frederique2021PlayfulEngagement}, as formulated by Gaver \cite{Gaver2004TheEngagement}, has been influential. There are elements of ludic design that seem to share characteristics with what has been done in relation to larp within the HCI community: playful ways of using design to explore concepts and sparking discussions. Drawing from, amongst other, examples of larp, the CHI'20 publication \textit{Technology for Situated and Emergent Play: A Bridging Concept and Design Agenda} \cite{AltarribaBertran2020TechnologyAgenda} specifically proposes a design agenda for the types of situated and emergent play that occur in larp. Exploring play and playfulness ties into research traditions much older than HCI, tapping into fields such as sociology and education, with for instance Huizinga \cite{Huizinga1955}, Piaget \cite{Piaget2013PlayChildhood}, Vygotsky \cite{Vygotsky1967PlayChild} and Caillois \cite{Caillois2001ManGames} as key references. Earlier work is often only very briefly touched upon in the publications, if at all. The Future Play'07 paper \textit{Pervasive Games in Ludic Society} \cite{Stenros2007PervasiveSociety} focuses especially on concepts of play and ludic, and looks at long-standing cultural trends influencing larp among other types of play, even though the paper focuses less on foundations in research traditions and looks more outside academia.

\subsubsection{Appropriation, co-creation and Participatory Design} This is another set of theories that has been identified in the publications surveyed. This ties into for instance Harrison and Dourish's theories on space and place \cite{Harrison1996Re-Place-IngSystems}, and also for instance \textit{Designing for Appropriation} \cite{Dix2007DesigningAppropriation}. A publication taking this as a focus is \textit{Design, appropriation and use of technology in larps} \cite{Segura2017DesignLarps}, presented at FDG '17. The TEI'21 publication \textit{Designing Gaming Wearables: From Participatory Design to Concept Creation} \cite{Jung2021DesigningCreation} is an example of a paper that explicitly focuses on larp and participatory design.

\subsubsection{Ethnography and ethnographic methods} Several publications engage with ethnography to some extent, although seldom explicitly. In the early ACE'06 publication \textit{Prosopopeia: experiences from a pervasive Larp} \cite{Jonsson2006Prosopopeia:Larp}, the authors describes how they did a type of 'ethnographic report' from a larp. Ethnographic methods and stances are not just HCI specific, but are used in relation to HCI, for instance what Müller describes as design ethnography \cite{Muller2021DesignEthnography}.

\subsection{Limitations}
As was mentioned, this overview is limited to only those articles in the ACM database explicitly mentioning larp. There are probably also more articles that uses larp-like methods without using the term. Further, there are many publications outside ACM, especially within the field of game studies \footnote{The Transformative Play Initiative, 
\url{https://www.speldesign.uu.se/research/games-and-society-lab/transformative-play/Transformative-Play-Initiative/}} and education \cite{geneuss_use_2021}, and there is even peer-reviewed journals; the International Journal of Role-playing \footnote{International Journal of Role-playing, \url{https://journals.uu.se/IJRP/index}}, and Japanese Journal of Analog Role-Playing Game Studies \footnote{Japanese Journal of Analog Role-playing Game Studies, \url{https://jarps.net/journal}},focusing on roleplays such as larp. This paper focuses specifically on larp in an HCI context, not on exploring larp or roleplay as a broader phenomenon. The literature review was conducted focusing on titles, abstracts and keywords. Statistics on affiliations and authors are based on ACMs meta-data. Overlapping authors who have changed affiliations etc. have not been tracked. 

\section{Larp Research Exemplars} \label{EXE}
In this section, we present larp research exemplars of larps that have been designed or studied within HCI or HCI-adjacent fields, to further contextualize for the reader the intersections between HCI and larp. All exemplars are research projects carried out by one ore several of the co-authors of this paper. This is a convenience sample from this author group and not meant to cover all types of larp research, rather it is meant to give some in-depth exemplars to create an understanding of the process of utilizing larp in HCI related research. For each exemplar there is an introduction, a description of the larp, and a description of how the exemplar relates to HCI topics and concepts, and how it has contributed to HCI related research.

\subsection{Larp: The Object}
This project has been included in numerous publications to-date. First, a CHI'20 paper Sensitizing Scenarios: Sensitizing Designer Teams to Theory \cite{Waern2020SensitizingTheory} focuses specifically on the use of those types of scenarios, among others The Object, and the experiences from developing and running the scenarios. A doctoral thesis, \textit{Articulating The User - a Discursive-Material Analysis of Humans in Interdisciplinary Design Collaborations} \cite{Rajkowska2022ArticulatingTheCollaborations}, was written within the project, the thesis among other things looks at how the scenarios helped design teams to articulate thoughts about the users. A book, \textit{Hybrid Museum Experiences - Theory and Design} \cite{Waern2022HybridDesign} was published at Amsterdam university press 2022, containing a chapter dedicated to the sensitizing scenarios, and containing ideas for how they can be used at museums.

The Object was designed within a research collaboration project together with artists, designers, museum professionals and researchers to support the creation of museum experiences. It was included in a collection of similar scenarios, all created to provide a sensitising and team-building experience for design teams working in a museum context, to reflect upon and learn about relevant concepts and theories related to their practice. The Object focuses on how we value objects, the life cycle of museum objects, and on what stories museums choose to tell. The script includes pre-workshop instructions, the larp itself, and suggestions for a structured debrief. The script is short but detailed, written to be accessible to design teams without access to expert larp facilitators. The Object was developed by three members of a research group, all with previous experience as larpers and larp designers. Basic facts about the larp can be found in Table \ref{tab:object}.

\begin{table}[h]
\caption{Basic facts about the larp: \textit{The Object}}
\label{tab:object}
\begin{tabular}{|l|l|}
\hline
\textbf{Name of larp}      & The Object                                                       \\ \hline
\textbf{Number of players} & 4-9 players, including facilitator                                        \\ \hline
\textbf{Venue}   & Ordinary room                                                             \\ \hline
\textbf{Duration}          & 1 hour 20 min                                                             \\ \hline
\textbf{Target group}      & Museum design teams                                                       \\ \hline
\textbf{Props}             & An object such as a small flower pot. Printed out roles and instructions. \\ \hline
\textbf{Short description} &
  \begin{tabular}[c]{@{}l@{}}Through small scenes the participants follow an object from its creation, \\ it being collected for a museum, it being exhibited and finally discarded \\ from the museum collection. The participants play different roles in \\ different scenes, for example villagers or museum personnel, and the \\ roles have different agendas and attitudes.\end{tabular} \\ \hline
\textbf{Purpose of the larp} &
  \begin{tabular}[c]{@{}l@{}}Sensitize design teams at museums towards concepts such as cultural \\ heritage, cultural appropriation and deaccession. Reflecting and \\discussing provenance, value, stories and life-cycles of museum \\ objects, what stories we tell at museums and what alternate stories \\ there could be. New museology, critical perspectives, experience design \\ and use of technology in cultural heritage practices.\end{tabular} \\ \hline
\textbf{Larp Designers}    & Karin Johansson, Jon Back, Annika Waern                                   \\ \hline
\end{tabular}
\end{table}

\subsubsection{HCI relevant game mechanics}
This scenario makes use of some common techniques from educational and entertainment larps. A central design choice is that the larp contains time jumps, to allow participants to follow a sequence of events and detach themselves from the current situation, which would typically be a design process with a looming deadline. Another important choice was to let participants change roles and perspectives, which allows the participants to step out of their own perspective and have an alibi for exploring and discussing attitudes and roles within the design team. The roles, especially those portraying museum personnel, were developed by observing museum design projects and working with museum professionals.

\subsubsection{Experience from running the larp}
The larp was designed in iterations, developed through observations and feedback from test groups consisting of designers and museum personnel. An early version was too time-consuming, therefore the Object was re-designed to be very short Furthermore, care was placed on creating a museum experience that felt more authentic, building on professional practices at museums.  \cite{Waern2020SensitizingTheory}. A researcher who observed when the scenario was run as part of a museum design process has described how it created group cohesion as well as reflection on the design situation at hand: ``\textit{Roleplaying proved both informative and silly and seemed to break down some social barriers to create a more relaxed atmosphere. One of the curators referred to it as therapy. The process finally enabled the partners to open up, articulating what had been problematic during the collaboration.}'' \cite{Rajkowska2022ArticulatingTheCollaborations} Museum personnel reflected on participating in the larp: ``\textit{For me, it felt really realistic...I don’t think that anyone is so clear about it, attitude of the perspective.}'' Another comment was, ``\textit{I think that since we often really have this sort of discussions in the museum, it’s good to try to see things from other side and make yourself create arguments for that.}''  A developer from the design company reflected on the usefulness of roleplay for design processes: \textit{``the main app designer reported on seeing the exercise particularly useful as a way to develop group cohesion and as a stepping stone to further design activities.''}

\subsubsection{Relation to HCI concepts and topics}
This larp is  HCI-relevant through its focus on tools for innovative experience design processes. While designed for a museum context, parts of it have been reused in a similar larp  for a wider scope of design situations at the edu-larp@CHI workshop at CHI'23 \cite{robinson_edu-larp_2023}. The work around the Object explicitly draws from participatory design and improvisational drama theories, and engages with concepts such as alibis for interaction, embodied experiences, authenticity, recontextualization and imaginative design explorations \cite{Waern2020SensitizingTheory}.

\subsection{Larp: The Deadline}
The design of the larps has been presented at Nordic DiGRA 2023, which had the theme of \textit{Interdisciplinary Embraces} \cite{Bowman2023EmployingCollaboration.}.

The Deadline is one of five scenarios developed for an interdisciplinary research curation center within a university. The center helps interdisciplinary research teams develop grant proposals and learn how to collaborate. The purpose of the scenario is to guide researchers through situations of escalated conflict that often arise during stressful situations. In the scenario, the characters are researchers in conflict related to whose research contributions are most relevant and valuable. The goal of the larp is to practice nonviolent communication skills such as “I-statements” within a professional environment, allow researchers to think beyond their positions and identify not just their interests, but also feelings and basic human needs underlying the conflict. Basic facts about the larp can be found in Table \ref{tab:deadline}.

\begin{table}[h]
\caption{Basic facts about the larp: \textit{The Deadline}}
\label{tab:deadline}
\begin{tabular}{|l|l|}
\hline
\textbf{Name of larp}      & The Deadline                                                       \\ \hline
\textbf{Number of players} & 3-5 + facilitator character                                        \\ \hline
\textbf{Venue}   & Ordinary room                                                             \\ \hline
\textbf{Duration}          & 3 hours                                                             \\ \hline
\textbf{Target group}      & Research scholars seeking funding in interdisciplinary collaborations                                                       \\ \hline
\textbf{Props}             & None \\ \hline
\textbf{Short description} &
  \begin{tabular}[c]{@{}l@{}}The scenario is set during the development process of a research grant \\ proposal. A group of multidisciplinary academics are trying to conduct \\ revisions on their proposal based on each of their interests and \\ research agendas. Conflicts emerge in the group that must be addressed \\ or the proposal will not meet the deadline. The purpose of the scenario \\ is to help researchers address the needs of everyone in a group most \\ peacefully and effectively through the practices of conflict \\ transformation \cite{Lederach2014TheTransformation, Gullick2020CONFLICTACADEMY, Rosenberg2015NonviolentLife} \end{tabular} \\ \hline
\textbf{Purpose of the larp} &
  \begin{tabular}[c]{@{}l@{}} 1) Practice receiving and responding to feedback that might be \\ upsetting or frustrating.\\ 2) Explain one’s research agenda in the hopes that others from different \\ disciplines can understand and value it. \\ 3) Persuade others to address one’s interests within a research group \\ through self-advocacy. \\ 4) Articulate interests, needs, and feelings through the use of \\nonviolent communication. \\ 5) Collaborate to search for win-win scenarios in order to complete \\ the proposal \\ successfully. \end{tabular} \\ \hline
\textbf{Larp Designers}    & Sarah Lynne Bowman, Josefin Westborg, and Kaya Toft Thejls                                 \\ \hline
\end{tabular}
\end{table}

\cite{Lederach2014TheTransformation, Gullick2020CONFLICTACADEMY, Rosenberg2015NonviolentLife}.

\subsubsection{HCI relevant game mechanics}
The larp has a technology team through the background and scope of the imagined research proposal. The group chooses from one of the four following “Big Idea” project pitches, which all are intended to fuse art and technology in some way to encourage social change:

UpCycle: Creation of an attractive app that helps people learn how to recycle in collaboration with local municipalities
Helping Hand: Community outreach program that hosts workshops blending arts and technology to empower at-risk youth
CyberCity: High tech art installations that incentivises co-creation with other residents throughout the city to improve connection, integration, and mobility.
ArtSmart: Research project that measures how digital arts education impacts interest in science and programming

Through this thematic scope, technology and the characters’ relationship to it could be considered another “character” in the game (Bienia 2016). The way this enters into the larp, however, varies with the knowledge of the players and their extrapolation of preferred research methods.

\subsubsection{Experience from running the larp}
Most runs of the larp focused primarily on the revision discussions, the conflicts within the group, and the internal conflicts of the characters. Players tend to deeply reflect on their own engagement in research groups during debriefs, rather than topics related to technology per se. An interesting insight from running the larp is how discussions of technology sometimes fueled feelings of superiority and/or inferiority within the group, where players playing the Technophile and Artist Practitioner characters sometimes will feel less secure in relation to the expertise of the other academics that are perceived as having higher academic status.

\subsubsection{Relation to HCI concepts and topics}
HCI is in many ways an interdisciplinary research subject, and HCI researchers will often find themselves in multi-disciplinary research groups.

\subsection{Larp: The Anywear Academy}
This larp was designed as part of a nationally funded project aimed at broadening interest and participation in STEM topics and careers. The purpose of this larp is to let middle-school girls familiarise themselves with technology design and development in ways that encourage building computational and design skills, interest, and self-efficacy. A paper published in DIS 2022 covers the concept, design, and impact of the first iteration of the larp \cite{Fey2022AnywearGirls}, while a series of publications are currently either under review or in progress. Additionally, the larp material is publically available for use \footnote{\url{https://www.anywear-academy.org/}}. Basic facts about the larp can be found in Table \ref{tab:aa}.
\begin{figure}
    \centering
    \includegraphics[width=.75\textwidth]{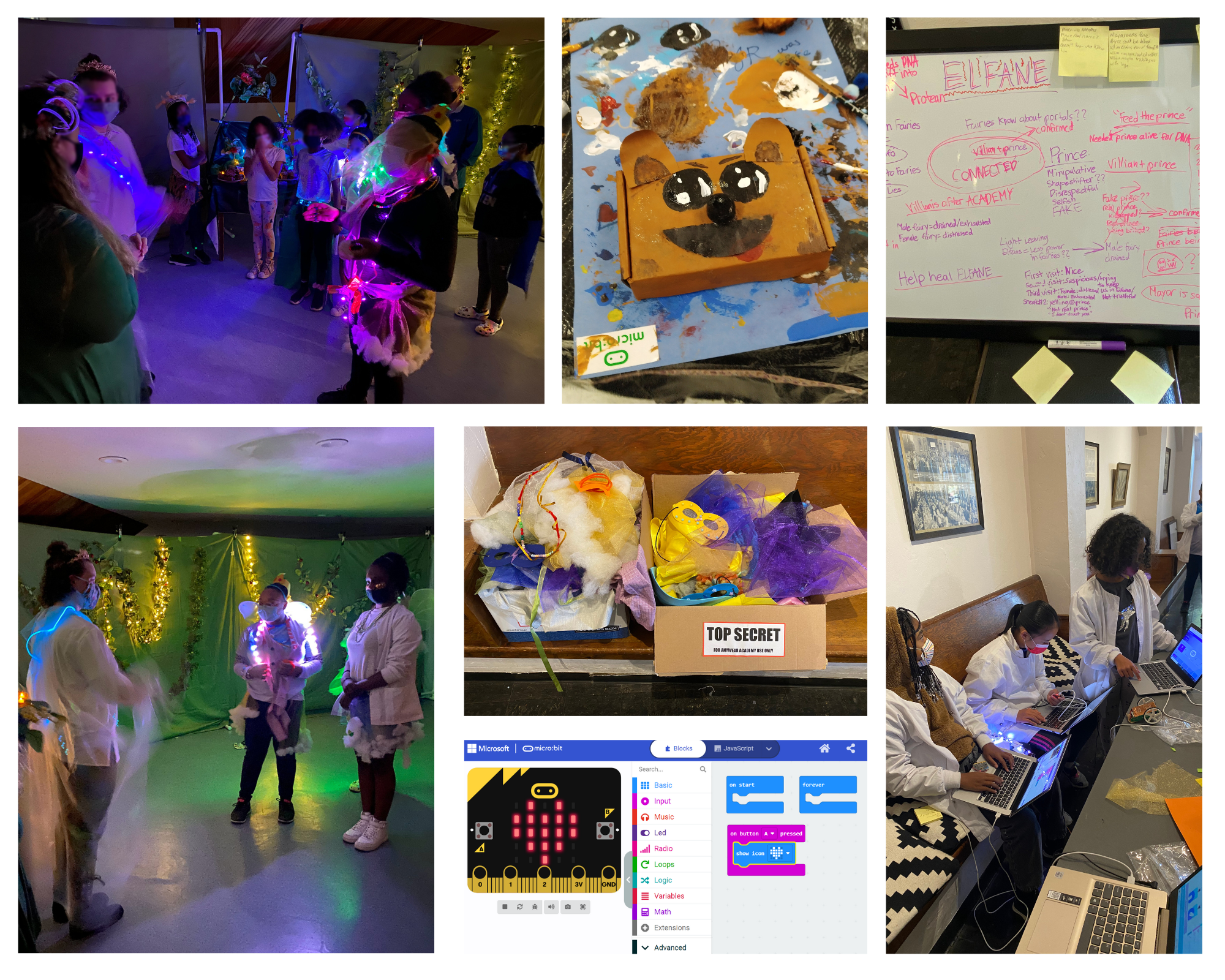}
    \caption{Photos from The Anywear Academy 5-day summer camp for middle school girls.}
    \label{fig:larp-camp}
\end{figure}

\begin{table}[h]
\caption{Basic facts about the larp: \textit{The Anywear Academy}}
\label{tab:aa}
\begin{tabular}{|l|l|}
\hline
\textbf{Name of larp}      & The Anywear Academy                                                       \\ \hline
\textbf{Number of players} & 7-15                            \\ \hline
\textbf{Venue}   & Multiple rooms, one for crafting/coding, and another for the missions                                                             \\ \hline
\textbf{Duration}          & 5-day camp                                                             \\ \hline
\textbf{Target group}      & Research scholars seeking funding in interdisciplinary collaborations                                                       \\ \hline
\textbf{Props}             & \begin{tabular}[c]{@{}l@{}} 1) Fairy-inspired items (crowns, vines tutus, fairy wings) \\ 2) Superhero-inspired items (capes, masks). \\ All costumes/wearables designed by campers. \end{tabular} \\ \hline
\textbf{Short description} &
  \begin{tabular}[c]{@{}l@{}} Campers participate in three different mission chains, or plots, \\ separated by theme and setting. Campers are trainees of the \\ `Anywear Academy', tasked with traveling undercover to different \\ worlds. There are three worlds the campers travel to: Elphame, a \\ fairy-themed magical world; Metro City, a superhero-themed world, \\ and Earhart station, a space station. Before every mission, an NPC \\ (non-player character–e.g. a camp staff member playing a role) \\ conducts a short briefing, explaining the mission objectives. Working \\ from this briefing, campers design wearables that allowed them to \\ blend into each setting, as well as meet specific technical challenges \\ laid out to them. \end{tabular} \\ \hline
\textbf{Purpose of the larp} &
  \begin{tabular}[c]{@{}l@{}} A larp-based summer camp in which middle school-age girls create \\ social wearables, toward building computational and design skills, \\ as well as interest in and self-efficacy about coding and design. \end{tabular} \\ \hline
\textbf{Larp Designers}    & \begin{tabular}[c]{@{}l@{}} James Fey, Katherine Isbister, Ella Dagan, The Game Academy, \\Raquel Robinson, Selin Ovali   \end{tabular}                              \\ \hline
\end{tabular}
\end{table}

\subsubsection{HCI relevant game mechanics}
Given the focus on learning technology and HCI concepts through immersive play, this larp creates the context for design exercises that facilitate learning HCI concepts. Key elements for social wearables design are covered through hands-on, iterative crafting and programming. The natural focus on group experience that is a part of larping supported the campers in designing technology to solve social problems (e.g., "create a costume to blend in with the fairy world that uses the LEDs to fool the fairies into thinking you have magic!"). Participants are taught embodied design methods to craft their own wearable technology. The technology used in the camp is the BBC:Microbit \cite{Austin2020TheWorld} with an attached strand of flexible LED lights (i.e., Adafruit Neopixel Dot Strands) for easy integration into their costumes (see Figure \ref{fig:larp-camp}). Over the course of the camp, the campers would alternate between courses on programming (in character as Anywear Academy trainees), crafting their wearables, and missions. The campers would receive a prompt to complete a wearable technology design that they would need to use in the upcoming mission (e.g., "Find a way to communicate nonverbally to one another with your costumes, as the space station systems are down and you won't be able to speak").

\subsubsection{Experience from running the larp}
By presenting a variety of methods of engaging with the educational material through open-ended crafting, design challenges, narrative justification, and collaborative play as valid, a number of participants who were otherwise uninterested in computer science learning became more interested on their own terms. Through a series of iterations, the tone of the material and the way it is presented to first-time larpers was tuned towards creating a more welcoming introduction to making and creating technology by highlighting the collaborative aspect of the experience over competition.
%%\subsubsection{HCI publications related to the larp}
%%A paper published in DIS 2022 covers the concept, design, and impact of the first iteration of the larp, while a series of publications are currently either under review or in progress. Additionally, the larp material is publically available for use.

\subsubsection{Relation to HCI concepts and topics}
This exemplar utilises larp as the context for teaching technical skills, toward broadening participation and interest in areas of HCI. Through experiences motivated by the larp narrative, participants gained a further understanding of designing technology for use in social contexts.

\subsection{Larp: True Colors Embedded in New Gyr}
This example is a Research-through-Design investigation of how to design wearables to support pro-social behavior. Academic researchers worked closely with a local larp community to design and deploy wearable technology in one instance of a larp series called New Gyr, exploring life in a fictional galaxy. The wearable was been demoed as part of UbiComp/ISWC'19 \cite{Dagan2019AVulnerability}. Key design research insights from this larp were presented as part of a CHI'19 paper \cite{Dagan2019DesigningVulnerability}. The work was also paved the way for intermediate-level knowledge \cite{Hook2012StrongResearch}, giving rise to a design framework for social wearables, and published at DIS'19 \cite{Dagan2019DesignWearables}, and in an MIT Book \cite{playful_wearablesMIT23}. Basic facts about the larp can be found in Table \ref{tab:newgyr}.

\begin{table}[h]
\caption{Basic facts about the larp: \textit{New Gyr}}
\label{tab:newgyr}
\begin{tabular}{|l|l|}
\hline
\textbf{Name of larp}      & New Gyr ('True Colors' tech was embedded within the larp)                                                       \\ \hline
\textbf{Number of players} & 109 (91 players; NPCs/staff: 15; 3 design researchers)                             \\ \hline
\textbf{Venue}   & \begin{tabular}[c]{@{}l@{}} Conference/retreat center with 3 buildings, multiple rooms and \\ outdoor spaces  \end{tabular}                                                           \\ \hline
\textbf{Duration}          & 3 days (approx. 32 hours of gameplay)                                \\ \hline
\textbf{Target group}      & General population >18 years old                                                       \\ \hline
\textbf{Props}             & \begin{tabular}[c]{@{}l@{}} Sci-Fi themed props and costumes, all designed and built by \\participants. Augment characters wore a wearable designed for \\ the purpose of the design research project. \end{tabular} \\ \hline
\textbf{Short description} &
  \begin{tabular}[c]{@{}l@{}} On a fictional planet several "human variants'' coexist, not always \\ in peace: Humans; Augments (Humans augmented with technology); \\ Evos (humans with altered genome); and Androids (artificial \\ intelligent agents). The Augments (13 players) were humans who \\ had resorted to technological augmentation to improve their \\ bodies, for practical reasons ``such as enhancing physical strength, \\expanding memory, or removing the need to breathe.'' Their \\ cultural status was low-class, taking on physical labor or working \\ in hazardous environments. The research group worked together \\ with the larp designers to develop digital wearables for the \\Augments,  designed to foster pro-social behaviour within the group \\ and with characters from the other factions.  \end{tabular} \\ \hline
\textbf{Purpose of the larp} &
  \begin{tabular}[c]{@{}l@{}} The larp was meant to be a sandbox for players to explore and \\experiment, with minimal organizer-introduced plot elements, \\rich in player-driven and player-organized plot(s). \end{tabular} \\ \hline
\textbf{Larp (Tech) Designers}    & Ella Dagan, Katherine Isbister, Elena Marquez Segura, James Fey                                 \\ \hline
\end{tabular}
\end{table}

\subsubsection{HCI relevant game mechanics} The wearables designed for the Augment characters implemented many mechanics, three of which were particularly relevant for the goals of encouraging pro-social behaviour. \textit{Overloads}: Narratively, the overloads represented problems experienced by the Augment, who would build ``electrical charge" which at some point would be violently discharged into their biological system, sending the Augment into a stunned state. Technically, the overloads were programmed into the wearable worn by the Augments, and signalled by the wearable flashing red. (There was no actual electricity discharged, so the player needed to roleplay the effect.) Overloads could be ended by \textit{Social Comfort and Healing}: Other characters could mitigate the pain and decrease the crisis duration through patting capacitive sensors at the back of the wearable, which was draped over the Augement's shoulders. Overloading was partly controlled by players, in that they would periodically happen automatically, but could also be triggered by the wearer to create a cool scene; or by another player with special triggering codes.  \textit{Stuns:} Augments could also use their wearables to inflict an electrical energy discharge upon another player, which would temporally stun them in a similar way. These stuns were also signalled through audiovisual effects. 

\subsubsection{Experience from running the larp}
The True Colors wearable enriched the larp experience. It was perceived as ``interactable '' and ``real'' facilitating immersion, scene building, and "increased possibilities for (social) interactions'' and interesting social experiences. The research showed that players did not use the stun function much nor hacking or inflicting overloads to other players. Instead, they enjoyed the overloads and healing functions, embracing ``vulnerability over power and control as resources to co-create interesting co-located social interactions. Far from being perceived as a burden to players, vulnerability was understood a source for meaningful personal and social experiences'' 

The research analyzed key design aspects and considerations to support designing for vulnerability, such as supporting ``big feels,'' supporting authentic self-presentation, overcoming difficulties together, and carefully working with information, choice and consent. 

\subsubsection{Relation to HCI concepts and topics} 
This larp is an example of a design research project conducted in the context of a leisure larp. \textit{Futuring} is at the core of this work, since the technology designed is meant to be used to better understand the emotional, social, and cultural impact of a future technology displaying sensitive information, such as affiliation, status, and health. \textit{Embodied Interaction} was a key consideration throughout the design process. The design was meant to "feel good on the body, and to affect and be affected by multiple senses (vision, hearing, and touch)'' \cite{Dagan2019DesigningVulnerability} as well as allow participants to``reflect and leverage physical and social bodily practices, such as social touch for comfort, and bonding.'' This was done through designing social affordances \cite{MarquezSegura2018DesigningDesigners}, and interdependency \cite{Isbister2017}. The wearables were co-designed with participants and larp organisers, and involved  embodied design methods \cite{MarquezSegura:2016:ES:2858036.2858486, Vega-Cebrian2023DesignCharacterization, Segura2016}, including a sensitizing and a bodystorming \cite{MarquezSegura:2016:ES:2858036.2858486, Segura2016} session in the larp site.

\section{Discussion and Synthesis} \label{discussion}
This papers aims to synthesise knowledge on larp in relation to HCI. The previous sections bring together knowledge in three distinct categories: a practitioner overview, a literature review and larp research exemplars. In this section, the content from the previous sections is discussed and further synthesised, with focus on the relevance for the HCI community in terms of future work and research. This is done through structuring and defining different aspects of the relationship between larp and HCI, which have been shaped via group discussions as described in the methods section, drawing upon the collective experience of the expert author group. 

Firstly, three different approaches to HCI related larp research are proposed; research into, for and through larp. Further a synthesised knowledge of the benefits of utilising larp in relation to HCI is presented, followed by identified challenges, and how those challenges could be handled. Together the discussion and synthesising sections provide inspirational guidance for how to utilise larp in the HCI community, as well as tentative suggestions of possible pitfalls, how to avoid these, and what could be an embryo of creating a best research practice connected to larp in relation to HCI research. Limitations to this study are discussed and future research suggested. 

As can be seen from the literature review as well as the larp research exemplars shared, there are many compelling reasons for HCI researchers to engage with larp. It may be useful to structure these along the lines of how Frayling \cite{Frayling1993ResearchDesign} characterised research's relation between art and design, as research \textit{into} larp, research \textit{for} larp, and research \textit{through} larp. The literature overview presented in this article demonstrates that even though research already has been done within all of these foci, there remain distinct gaps in the academic understanding and use of larp in the HCI context. 

\subsubsection{Research into larps}
Larp is a cultural expression in its own right. It is also a global phenomenon with many million players all over the world, mainly grassroot organized via social media and other digital platforms. Tech has increasingly been integrated in different ways into these larps, as has been presented in the practitioner overview. Despite all this, little has been published, as is pointed out in the literature review, within HCI research about existing larps practices. And when larps are studied, the research focus has largely been on experiments with innovative technology, rather than established practices of technology use in larp, that the practitioner overview shows is abundant. This also goes for larp as relevant interaction design practice. The consequence is that we know too little about how larps are designed, and what considerations are important for the use of technology within them. Understanding larp design better could inform other types of experience and/or activity design, especially in relation to co-creation, first person perspectives and in the creation of experiences that are at the same time individualized and collective, given the social and collective nature of larp experiences. An example of a research project engaging with research into larp could be a comparative study between an online larp and an on-site larp, to identify what interactions are shared between the two types of larp, and to what extent for instance the embodiment and tangibility of a physical version of a larp affects the experience. 

\subsubsection{Research for larps}
Hand-crafted novel technologies and existing digital tools are already used in larps, as is shown in the practitioner overview, and larp designers and players use practical knowledge to build larps, costumes, and settings. This practice remains under-researched, and could potentially inform design in adjacent domains. Furthermore, many larp designs require facilitators to be physically present and active at all times, which is why many larps become confined to specific spaces or small numbers of players. Technology solutions could support such larps in scaling up or becoming more easily re-stageable. In the literature review, this type of research seems scarce, even though, as shown previously in this paper, larp is a vibrant and large subculture so there could be interesting to support and inform larp design processes, and to for instance find technology solutions that could be key in supporting a wider use of larp for serious purposes such as training or therapy. 
Finally, even though technology-related HCI research has been dominated by technology innovation, there is still room for innovation in the use of technology for creating new larp experiences. An example of a research project engaging with research for larp could be a Research through Design project developing a diegetic magical looking glass through which different larp groups could gain secret knowledge about other groups, and then try and tweak the design to create maximum leverage from a larp interaction perspective, identifying which features of such a design that would best promote immersion and a better larp experience.

\subsubsection{Research through larps}
Several publications have already utilised larp as a tool to explore relevant HCI topics, as is shown in the literature review and through the larp research exemplars. One reason for this is the versatile format that larp offers. In particular, larp has a role as a design tool that can be used throughout the whole design process \cite{MarquezSegura2019LarpingMethod} to sensitize designers and other stakeholders to particularly sensitive topics, situations, and scenarios; to come up with interesting technology design ideas (e.g. \cite{Fey2022AnywearGirls}); and to test, evaluate and even iterate technology designs and prototypes (e.g. \cite{MarquezSegura2018DesigningDesigners, Dagan2019DesigningVulnerability}). Larps also align well with critical approaches such as design fiction; they are particularly well suited to imagine futures and future designs and critically experience and reflect on them and their impact. Futuring was identified as a reoccurring topic in both the literature review, as well as seen occurring though the practitioner overview. Redström, in his book \textit{Making Design Theory}, argues that design research could and should engage with futuring: "Design research addressing matters such as attitudes, values, and behaviours, does not necessarily depend on future technologies or opportunities to be possible. On the contrary, it depends on questioning what is now, asking what if things were already different" \cite{Redstrom2017MakingTheory}. Larp, as mentioned in this paper, is about making a mental journey, putting oneself in an immersive situation and through transformative play imagining different realities. This gives unique opportunities to engage in futuring, independent of future technologies, since we can pretend through the larp that they already exist, and then explore what that would mean for a group of people, part of the society, or for our own behaviors and emotions. How this can be done can be seen for instance by the True Color exemplar. Further, larps can be suitable when working with value-based design research, such as value sensitive design, socially responsible design, reflective design, and design for social sustainability, as it is a good way of shifting perspectives, sparking discussions and getting emotional understanding for different concepts and situations. Larps could also be utilized in Research through Design, where one might normally use design ideation methods such as Wizard of Oz \cite{Kelley2018WizardJourney}, rapid prototyping, use case theatre \cite{Schleicher2010, Oulasvirta2003UnderstandingBodystorming}, or object theater - stakeholder drama \cite{Wilde2017}. Larp could be used instead for a deeper first person perspective to try out designs by enacting real world user situations, by roleplaying user reactions and interactions. A key difference between these aforementioned design methods and larp is that larp places a stronger focus toward the emotional, social, and emergent experiences of the participants, rather than having a set outcome or objective. An example of a research project engaging with research through larp could be to create and utilize a sci-fi larp set in a technology-dominated smart city, to gain HCI relevant understanding of how it could be like to live in such a city, using larp as a tool for futuring and speculative design. 

\subsection{Benefits of Utilising Larp in HCI}
The expert group, having engaged much with larp in relation to HCI topics and research, has brought together knowledge on and discussed the benefits of utilizing larp. Synthesising some of the themes uncovered, in the previous sections of this paper and in those discussions, several possible benefits of utilising larp in relation to HCI have been identified:

\begin{itemize}
    \item \textbf{Versatility} - Through world building and players agreeing on the narrative, larp can be easy to adapt to different situations and topics.
    \item \textbf{Explorations} - Being based in imagination, larp can afford exploration in regards to ourselves and our world, but also other worlds, fictions and roles, through imaginative playfulness.
    \item \textbf{Shifting perspectives} - The role-taking in larp can be utilised to understand others' points of view, or to empower and see possibilities.
    \item \textbf{Transformative and emotional qualities} - Through shifting perspectives and the design of experiences, as well as through processing these in post-game workshops, it seems larps could be used for creating experiences that elicit change in players.
    \item \textbf{Embodiment} - Larp invites the use of all senses to create their experiences, unlike for instance tabletop rpgs.
    \item \textbf{Futuring} - Larp offers the chance to readily create an immersive environment within which to experiment with e.g. future technologies, using the larpers' own collective imagination - making those futures possible for a short while, within the story told.
    \item \textbf{Playification} - Larp offers many of the benefits of gamification but without strong extrinsic motivations such as badges, scores and leader-boards. Instead, larp motivates through playful exploration and co-creative storytelling.
\end{itemize}

\subsection{Challenges with Utilising Larp in HCI Research}
Despite all these opportunities, there also comes challenges with utilising larp in HCI research. These challenges need to be thought through and reflected upon when making informed decisions as to if and how larp should be utilised. Some of those challenges are more general when designing and facilitating larps, and there is already an abundance of literature on this \cite{2019LarpExperiences, Back2014TheLarp, DuusHenriksen2011ThinkKP2011}, but there are also specific challenges for utilising larp in HCI relevant research, and this is the focus of this section.

\textbf{Special skills required }- Designing and running a larp is no easy task. It requires skills related specifically to larp, including general knowledge of various common larp rules (which vary locally),  and also taps into a range of more general skills including designing, facilitating, improvising, acting skills, practical planning and much more. These are competencies that larp designers hone over a long time through successes and failures, and can be hard to attain as a researcher. Hence, at least when a larger production is at hand, researchers typically benefit from actively involving and collaborating with a larp design and production team, who can assume the primary responsibility for larp design and facilitation. In such collaborations, it is important to make clear the role of the researchers, who may for example take the lead on technology design or on studying a production (e.g. \cite{MarquezSegura2018DesigningDesigners, Dagan2019DesigningVulnerability}.

\textbf{Demanding activity} - Larp demands a lot not only from the organizers but also from participants. It takes time to prepare as well as participate, and participation drains mental energy. A consequence of the large burden that larp places on participants, it that it is not always easy to get people to sign up, even for entertainment larp, and even more so for experimental and research purposes. Designs must offer opt out possibilities and a well thought through on-boarding process, to help participants engage in the intended way. This challenge can be addressed by regulating the length of the experience, avoiding rule-heavy designs, and avoiding emotionally heavy topics when these are not in focus.

\textbf{Ethical challenges} - Research in relation to larp raises a number of ethical considerations. Participation can be very emotional, and thus also potentially triggering. Further, larp does not only affect the players' minds but also their bodies, and there are real physical risks related to larp participation. The improvised and co-created format makes it hard to foresee every ethical problem that can arise, forcing facilitators to use a sort of ethical-considerations-on-the-fly-approach that can be hard to navigate as a researcher. International collaborative research projects utilising larp can potentially become a bureaucratic nightmare of different laws, regulations, rules and ethical approval processes. 

\textbf{Documentation issues -} Documenting larp for research purposes is tricky. Larp happens in the moment, and much of it in our heads as individual experiences. As researchers we have found that photos, written documentations, videos and such do not fully capture the larp experience. When larps are run for research purposes, it is imperative to use a host of documentation methods in parallel, both during and after the larp. A combination of documenting game documents and props, film and photo, and gathering stories from the participant and organiser perspectives both during and after the larp has proven effective in previous work. However, this approach also leads to a difficult and time-consuming qualitative analysis process.

\textbf{Rerunability }- Larp experiences are not necessarily repeatable due to the large amount of co-creation involved. Even when run from a stable and detailed script, larp events, and for that reason experiences, can vary wildly between runs, or even between players in the same run. While this is something that players typically appreciate, it also means that some types of research - such as the stringent study of the effect of specific design choices - are difficult to do by running singular larps.

The versatile, co-creational and improvisational character of larp is also what makes it a uniquely interesting art form. Larp can be partially "railroaded" (controlled by designers and facilitators), but there is limit if they are to remain larps. Larps are emergent, and as Gaver points out, emergence could and should be a vital part of design research \cite{Gaver2022EmergenceResearch}.

\subsection{Handling the Challenges}
There is a need to develop robust methods and research practices that are adapted to the design logic of larps. In this, it is vital to learn from fields such as education research, ethnography and anthropology, on how to stage and study situated experiences.

For researchers who wish to utilise larp as a research tool, there is a need for HCI-relevant training in designing and staging larps. In this, much can be learned from larp practitioners, for instance regarding safety practices \cite{Koljonen2020LarpFundamentals} and the practicalities of designing and facilitating larp \cite{2019LarpExperiences}. It is also important to understand when larp can be relevant, and when it is not the best tool.

We propose to facilitate more discussions within the HCI community to find new methods for documentation, and develop formats for rigorous mediation of research in relation to larp. There are several practical examples available in the literature that can inform how to structure methods and formats of larps \cite{Waern2020SensitizingTheory, Fey2022AnywearGirls}, but future research focusing specifically on developing such methods could be helpful for the research community. This discussion needs to also include how design research focused on larp can and should be presented, as the paper format can be ill-suited for situated and embodied experiences such as larp. 

\subsection{Limitations and Future Research}
Inevitably, the set of authors who have crafted this synthesis have shaped and limited the contours of the results, biasing what is written to contexts within which the authors have more experience and knowledge. This bias includes emphasis on Nordic countries and the US. Engaging larpers and larp researchers from other global communities in future work would broaden exemplars and insights. In addition, a deeper literature review that ventures beyond larp as a keyword could reveal a much larger swath of HCI work that takes inspiration from and speaks to larp. 

In addition, we acknowledge that much of the literature and larp research exemplars highlighted in this paper are particularly utilising \textit{Nordic} style larps. We see this as a gap that is worth exploring further, as an opportunity for more research in HCI to focus on exploring larps and how to apply HCI topics to larps outside this style - potentially Jubensha or American-style larp. Further the larp research exemplars are all variations of research through larp, this is also the predominant type of larp related research in HCI that has been conducted the latest years, as can been seen through the literature review. Future research could include investigating the other types of larp related research in HCI, that are now identified and explicated in this paper.

In doing the work of synthesising past research on larp within the HCI community, we have raised possibilities in each section for possible future research, for example, engaging more closely with existing larp communities to understand their technology use and to design technological support for their practices, or focusing more closely on the use of larp in formal education, and/or children as a target use community. In general, we encourage the HCI community to engage more actively with the larp community and affiliated scholarly communities focused on larp, toward generating valuable and fruitful HCI insights and prototypes. 

Synthesising knowledge on larp in relation to HCI, as has been done in this paper, can create a platform for future research. The paper has brought in new knowledge on what topics and themes have gained focus on larp related research in HCI, from the first larp related publications in ACM up until today, and also identifies less and potentially under researched topics. Further this paper provides a new synthesised mapping of terminologies and approaches focused on larp in relation to HCI, inside and outside academia. Further the gathering of a comprehensive collection of examplars and annotating them can provide inspiration, method ideas and formats for future researchers. Together with the comprehensive lists of benefits and challenges of utilising larp in relation to HCI research, the content of this paper can help facilitating, improving and shaping future research projects utilising larps in relation to HCI.

\section{Conclusion}
In this paper we highlight the varied use of larp as a global phenomenon and form of cultural expression, and point out that it is a unique and interesting type of interaction design to study in relation to HCI. We discuss the many ways that larp engages with technology, digital aspects and futuring, relating to topics relevant to the field of HCI. We present different ways that larp has already been utilised in relation to HCI research, considering research into, for and through larp. Despite what has already been done, there are several missed opportunities, and in this paper we propose ways of tapping into those potentials when it comes to larp in relation to HCI research. 

Within the HCI community there is a lot of relevant knowledge when it comes to interaction design, experience design, co-creative design processes, playful interaction and much more that could be relevant to research for improving, enhancing and developing larp and larp-like methods. Further, larp can be of use as a tool to conduct research, and larp and larp-like methods can be used to explore a wide range of HCI relevant topics, especially when it comes to value based design and different types of futuring.

To facilitate this kind of progress, we believe an important step forward is to find a common foothold. Definitions of what larp is and increased meaningful academic conversion has already started in some of the workshops and events mentioned. We see this paper as a step toward expanding the HCI community's foundational knowledge about larp, and also, contextualizing existing work at the intersection of larp and HCI, toward generating strong future research. 

In short:
\begin{itemize}
    \item Larp is for several reasons relevant for the HCI community
    \item It is possible to utilise larp to make interesting research contributions
    \item We propose to find more ways to tap the potential of larp more deeply in relation to HCI research
    \item We present a consolidation of current streams of work, and invite more academic dialog focused on larp
    \item We see benefit in bridging design work done in industry and the larp communities, with design knowledge within the field of HCI.
\end{itemize}

We look forward to more academic dialogue centered on larp in relation to HCI, and we are eager to explore how to deepen the relationship between larp and the HCI community. This paper and the synthesizing done can help inform future research, as well as bring a coherent knowledge frame of larp into the HCI research community.

\bibliographystyle{ACM-Reference-Format}
\bibliography{whylarp2, references}
%\bibliography{references.bib}

\appendix
\section{Table of Larps}
\label{appendix: larps}
\begin{landscape}
\begin{table}[]
\caption{Table of larps mentioned in the paper. The information is gathered from homepages, and in some cases personal experiences from having attended the larps}
\resizebox{\columnwidth}{!}{%
\begin{tabular}{@{}|l|l|l|l|l|l|l|l|@{}}
\toprule
\textbf{Number} &
  \textbf{Name of larp} &
  \textbf{Genre/ Type} &
  \textbf{Short description} &
  \textbf{Year} &
  \textbf{Organizers} &
  \textbf{Country} &
  \textbf{External Link} \\ \midrule
1 &
  {\color[HTML]{495365} Alkemistens verkstad} &
  Online larp/ urban fantasy &
  \begin{tabular}[c]{@{}l@{}}An online larp where participants play students at a magical school, \\ learning alchemistry, connected by magical screens.\end{tabular} &
  2019 &
  LajvVersktaden &
  Sweden &
  {\color[HTML]{495365} \url{https://lajvverkstaden.se/}} \\ \midrule
2 &
  {\color[HTML]{495365} Ancient hours} &
  VR larp &
  \begin{tabular}[c]{@{}l@{}}VR-larp. The story of Two demi-gods, Love and Memory, held captive \\ by humans, tortured and pressured into giving up some of their powers. \\ Conjuring their joint abilities, however, they manage to meet in their \\ common memories.\end{tabular} &
  2019 &
  {\color[HTML]{495365} Joffe Rydberg} &
  Sweden &
  {\color[HTML]{495365} \url{https://alexandria.dk/en/data?scenarie=14058}} \\ \midrule
3 &
  {\color[HTML]{495365} Android Beginning} &
  Sci-fi &
  \begin{tabular}[c]{@{}l@{}}Bladerunner inspired, philosophical on humanity and androids. \\ Specially built indoor venue. Nordic larp style, 360°\end{tabular} &
  2022 (campaign) &
  Atropos. Julia Woods. &
  Sweden &
  {\color[HTML]{495365} \url{https://beginnings.atropos.se/}} \\ \midrule
4 &
  {\color[HTML]{495365} Aurum LARP} &
  Steampunk/ fantasy &
  \cellcolor[HTML]{FAFBFC}\begin{tabular}[c]{@{}l@{}}Neo-Victorian Steampunk larp, with dying elf magic, new frontiers \\ and power-hungry syndicates.\end{tabular} &
  2023 &
  Aurum larp &
  USA &
  {\color[HTML]{495365} \url{https://scifixfantasy.com/events-all/aurum-larp/}} \\ \midrule
5 &
  {\color[HTML]{495365} Belvedere} &
  History/ literature larp &
  \cellcolor[HTML]{FAFBFC}\begin{tabular}[c]{@{}l@{}}Set in an alternative 17th century France, inspired by Three musketeers \\ and Da Vinci. Technology, prophecies, religion and politics.\end{tabular} &
  2023 &
  RSV Arcana &
  Netherlands &
  {\color[HTML]{495365} \url{https://www.arcana.nl/evenementen/belvedere/}} \\ \midrule
6 &
  {\color[HTML]{495365} Beyond the Neural Horizon} &
  Sci-fi &
  \begin{tabular}[c]{@{}l@{}}Hackers, where the internet is physically manifested in the rooms. \\ Specially built indoor larp venue. Nordic larp style, meta techniques.\end{tabular} &
  2019 &
  Atropos &
  Sweden &
  {\color[HTML]{495365} \url{https://alexandria.dk/en/data?scenarie=9900}} \\ \midrule
7 &
  {\color[HTML]{495365} Blodsband Reloded} &
  Wasteland &
  \cellcolor[HTML]{FFFFFF}\begin{tabular}[c]{@{}l@{}}Mad max inspired larp, set in a crazy wasteland future. \\ Venue: a big sandpit, old industry buildings, and city built of cargo.\end{tabular} &
  2018 (and reruns) &
  {Blodsband Reloded} &
  Sweden &
  {\color[HTML]{495365} \url{https://lajvhistoria.se/lajv/Blodsband\_Reloaded\_23.1502}} \\ \midrule
8 &
  {\color[HTML]{495365} Fight like a girl} &
  Cyberpunk &
  Dystopian feminist gender war larp &
  2018 &
  Frida Gamero, Susanne Vejdemo &
  Sweden &
  {\color[HTML]{495365} \url{https://flaglarp.wordpress.com/vision/}} \\ \midrule
9 &
  {\color[HTML]{495365} \begin{tabular}[c]{@{}l@{}}Fortune \&amp; Felicity: \\ A Jane Austen Larp\end{tabular}} &
  History/ literature larp &
  \begin{tabular}[c]{@{}l@{}}Jane Austen inspired, on the drama between passionate love, \\ necessity of money and the silliness of people. 360° Nordic larp style.\end{tabular} &
  2017 &
  Anna Westerling, Anders Hultman. &
  Sweden &
  {\color[HTML]{495365} \url{https://anna905.wixsite.com/austenlarp}} \\ \midrule
10 &
  {\color[HTML]{495365} Greylight 2142} &
  Cyberpunk &
  \begin{tabular}[c]{@{}l@{}}A Cyberpunk world of greed, oppression, and dystopian future. \\ Political. Indoor venue.\end{tabular} &
  2023 &
  Greylight &
  Germany &
  {\color[HTML]{495365} \url{https://greylight.de/}} \\ \midrule
11 &
  {\color[HTML]{495365} Havenhollow} &
  Medieval fantasy &
  Medieval Fantasy, battles and rules. Campaign larp. &
  2015 (reruns) &
  Underworld &
  Japan &
  {\color[HTML]{495365} \url{https://www.facebook.com/groups/1520452058215307/}} \\ \midrule
12 &
  {\color[HTML]{495365} Insomnia larp} &
  Horror &
  Horror larp of investigation of the unknown. &
  2016 (and reruns) &
  Insomnia LRP &
  UK &
  {\color[HTML]{495365} \url{https://www.facebook.com/insomnialrp/}} \\ \midrule
13 &
  {\color[HTML]{495365} Lovestories by ABBA} &
  Musical larp &
  \begin{tabular}[c]{@{}l@{}}The players play a band during their last summer tour in the \\ late 1970s. Focus on love, drama, music and parties. \\ Emotions are displayed through songs. Chamber/ blackbox larp.\end{tabular} &
  2022 &
  Anna Westerling &
  Sweden &
  {\color[HTML]{495365} \url{https://scenariofestival.se/archive/scenarios-2022/lovestories-by-abba/s-2022/lovestories-by-abba/}} \\ \midrule
14 &
  {\color[HTML]{495365} Mutanternas tidsålder} &
  Sci-fi &
  \begin{tabular}[c]{@{}l@{}}Childrens summercamp larp about mutants and humans. \\ Daily larps over a period of 6 weeks\end{tabular} &
  2016 (and reruns) &
  LajvVerkstaden &
  Sweden &
  {\color[HTML]{495365} \url{https://lajvverkstaden.se/}} \\ \midrule
15 &
  {\color[HTML]{495365} Mythodea} &
  Fantasy &
  \begin{tabular}[c]{@{}l@{}}Huge outdoor fantasy battle larp, around 10 000 participants per larp. \\ American larp style.\end{tabular} &
  2004 (yearly) &
  Mythodea &
  Germany &
  {\color[HTML]{495365} \url{https://mythodea.de/en/home-en/}} \\ \midrule
16 &
  {\color[HTML]{495365} Nittonhundra} &
  Steampunk &
  A steampunk science conference at the turn of the century &
  2015 &
  Sara Örn Ternstrand &
  Sweden &
  {\color[HTML]{495365} \url{https://steampunkisverige.wordpress.com/category/lajv/}} \\ \midrule
17 &
  {\color[HTML]{495365} Omgiven av idioter} &
  Sci-fi/ Convention larp &
  \begin{tabular}[c]{@{}l@{}}Homoristic/dystopic larp on social media addiction. \\ Chamber/ blackbox larp.\end{tabular} &
  2019 &
  Jon Back. Karin Johansson. &
  Sweden &
  {\color[HTML]{495365} \url{https://lajvkonvent.files.wordpress.com/2020/10/broschyr.pdf}} \\ \midrule
18 &
  {\color[HTML]{495365} Terra Incognita} &
  Lovecraftian horror/ 1930s &
  \begin{tabular}[c]{@{}l@{}}Lovecraftian horror set in the 1930s. \\ Played in a village and forest. Nordic larp style, 360°\end{tabular} &
  2013 &
  \begin{tabular}[c]{@{}l@{}}Berättelsefrämjandet\\ Olle Nyman\\ Sebastian Utbult\end{tabular} &
  Sweden &
  {\color[HTML]{495365} \url{https://lajvhistoria.se/lajv/Terra\_Incognita.45}} \\ \midrule
19 &
  {\color[HTML]{495365} \begin{tabular}[c]{@{}l@{}}The Austen Experience, \\ A Winter’s Ball\end{tabular}} &
  \begin{tabular}[c]{@{}l@{}}History/Litterature larp. \\ Cyberpunk, Fusion\end{tabular} &
  \cellcolor[HTML]{FFFFFF}{\color[HTML]{252525} \begin{tabular}[c]{@{}l@{}}Inspired by the works of Jane Austen as well as the tv-show \\ Westworld, and takes place at a Company-owned Austen theme park \\ with AI units and Androids. Location: a castle. Nordic larp style, 360°\end{tabular}} &
   &
  Atropos &
  Sweden &
  {\color[HTML]{495365} \url{https://austen.atropos.se/}} \\ \midrule
20 &
  {\color[HTML]{495365} The Epic Western Larp,} &
  Western &
  Big wild west larp, outdoors and village. 360° &
  2025 &
  Ultimate Western &
  Spain &
  {\color[HTML]{495365} \url{http://www.ultimewestern.com/en/accueil-en/}} \\ \midrule
21 &
  {\color[HTML]{495365} Turings fråga} &
  Future &
  \begin{tabular}[c]{@{}l@{}}Philosophical larp set in a non-defined future, were participants try \\ to pass a test to enter a prestigious academy, by identifying the android \\ in the group, using different criteria of what it means to be human. \\ The larp was run in the Arch-Cathedral of Sweden. Nordic larp style.\end{tabular} &
  2013 &
  {\begin{tabular}[c]{@{}l@{}}Frida Gamero\\ Sara Engström\\ Johan Dahlberg,\\ in collaboration with the \\ Swedish Church.\end{tabular}} &
  Sweden &
  {\color[HTML]{495365} \url{https://lajvhistoria.se/lajv/Turings\_Fraga.391}} \\ \midrule
22 &
  {\color[HTML]{495365} Voidship Concordia} &
  Sci-fi &
  Warhammer 40k -inspired chamber-larp with gamistic elements &
  2015 &
  Karl Bergström &
  Sweden &
  {\color[HTML]{495365} \url{https://spelkult.se/kalender/event/voidship-concordia/}} \\ \midrule
23 &
  {\color[HTML]{495365} Zero} &
  Sci-fi &
  \begin{tabular}[c]{@{}l@{}}Players play a crew aboard a spaceship, leaving the dying earth behind. \\ A technology enhanced spaceship venue. Futuring, political, \\ Nordic larp style, 360°\end{tabular} &
  2023 &
  Not Only Larp &
  Spain &
  {\color[HTML]{495365} \url{https://zero.notonlylarp.com/}} \\ \midrule 
24  &
  {\color[HTML]{495365} Last Will} &
  Dystopian Future &
  \begin{tabular}[c]{@{}l@{}}A larp aimed at looking at modern slavery through the lens of a fictional future. \\A fading human dignity in a world run by money and consumption in which \\ people can be bought and sold as commodities. \\ Nordic larp style, 360°, combined with meta techniques \end{tabular} &
  2014,2015 &
  {\begin{tabular}[c]{@{}l@{}}Ursula\\ Annica Strand\\ Frida Gamero\\ Sofia Stenler \end{tabular}} &
  Sweden &
  {\color[HTML]{495365} \url{https://www.foreningenursula.se/last-will/}} \\ \midrule 25 & {\color[HTML]{495365} Odysseus} &
  Sci-fi &
  \begin{tabular}[c]{@{}l@{}}Survivors from a destroyed space colony try to find the home of their ancestors. \end{tabular} &
  2019, 2024 &
  Illusiary & Finland &
  {\color[HTML]{495365} \url{https://www.odysseuslarp.com/}} \\ \midrule 26 & {\color[HTML]{495365} Mission Together} &
  Sci-fi &
  \begin{tabular}[c]{@{}l@{}}A space mission full of power struggles and cultural differences \\ between people from different planets. \end{tabular} &
  2020 &
  Not Only Larp &
  Spain &
  {\color[HTML]{495365} \url{https://notonlylarp.com/}} \\
  \bottomrule
\end{tabular}%
}
\end{table}
\end{landscape}



\end{document}